\documentclass[12pt]{JHEP3}
\usepackage{amsfonts} 
\usepackage{amssymb}
\usepackage{lscape}
\usepackage{cite}
\usepackage{epsfig}
\usepackage{multirow}
\usepackage{longtable}
\usepackage{amsmath}
\author{}

\newcommand{\be}{\begin{equation}}
\newcommand{\ee}{\end{equation}}
\newcommand{\ba}{\begin{array}}
\newcommand{\ea}{\end{array}}
\newcommand{\bea}{\begin{eqnarray}}
\newcommand{\eea}{\end{eqnarray}}

\def\IR{\relax{\rm I\kern-.18em R}}

\def\IP{\relax{\rm I\kern-.18em P}}
\def\inbar{\vrule height1.5ex width.4pt depth0pt}
\def\IC{\relax\,\hbox{$\inbar\kern-.3em{\rm C}$}}

\def\a{\alpha}

\def\K3{{\bf K3}}

\def\n2d{\cN_{V^*}^{\otimes 2}}

\def\IC{\mathbb{C}}

\def\IR{\mathbb{R}}

\def\IP{\mathbb{P}}

\def\cN{{\mathcal N}}

\def\nn{\nonumber}

\def\to{\rightarrow}

\title{Revisiting light stringy states \\ in view of the 750 GeV diphoton excess}
\author{
Pascal Anastasopoulos$^{1}$\footnote{pascal@hep.itp.tuwien.ac.at},~
Massimo Bianchi$^{2}$\footnote{massimo.bianchi@roma2.infn.it},~
\\
$^1$ Technische Universit\"at Wien \\ ~~ Institut f\"ur Theoretische Physik \\ ~~ A-1040 Vienna, Austria\\
$^2$ Dipartimento di Fisica and Sezione I.N.F.N. \\ ~~ Universit\`a di Roma ``Tor Vergata''
\\ ~~ Via della Ricerca Scientifica 1, 00133 Roma, Italy
}
\date{}

\maketitle
\abstract{We investigate light massive string states that appear at brane intersections. They replicate the massless spectrum in a richer fashion and may be parametrically lighter than standard Regge excitations. We identify the first few physical states and determine their BRST invariant vertex operators. In the supersymmetric case we reconstruct the super-multiplet structure. We then compute some simple interactions, such as the decay rate of a massive scalar or vector into two massless fermions. Finally we suggest an alternative interpretation of the 750 GeV diphoton excess at LHC in terms of a light massive string state, a replica of the Standard Model Higgs.}
\preprint{TUW-16-01\\ ROM2F/2016/01}
 \clearpage
\begin{document}

\newpage

\section{Introduction}

Starting from the pioneering works on the systematics of open string theories \cite{Sagnotti:1988uw, Bianchi:1988fr, Bianchi:1988ux, Pradisi:1988xd, Bianchi:1989du, Dai:1989ua, Bianchi:1990yu, Bianchi:1990tb, Bianchi:1991rd, Bianchi:1991eu}, that led to the first chiral model in $D=4$ \cite{Angelantonj:1996uy}, intersecting D-brane models \cite{Polchinski:1995mt, Berkooz:1996km} proved to be a promising class of models that allow to embed the Standard Model or extensions thereof in String Theory\footnote{For recent reviews on D-brane model building as well as specific MSSM D-brane constructions, see \cite{Blumenhagen:2005mu, Blumenhagen:2006ci, Marchesano:2007de, Cvetic:2011vz, Ibanez:2012zz} and references therein.}.

One of the most exciting properties of this class of models is that they allow for a very low string scale, even of order of a few TeV's \cite{ArkaniHamed:1998rs, Antoniadis:1997zg, Antoniadis:1998ig}, leading to interesting signatures detectable at particle colliders such as the LHC or in cosmological observations such as Planck. Particular attention has been devoted to anomalous $Z'$ physics (see, e.g. \cite{Kiritsis:2002aj, Antoniadis:2002cs, Ghilencea:2002da, Anastasopoulos:2003aj, Anastasopoulos:2004ga, Burikham:2004su, Coriano':2005js, Anastasopoulos:2006cz, Anastasopoulos:2008jt, Armillis:2008vp, Fucito:2008ai,  Anchordoqui:2011ag, Anchordoqui:2011eg, Anchordoqui:2012wt}), 
Kaluza Klein states (see, e.g. \cite{Dudas:1999gz, Accomando:1999sj, Cullen:2000ef, Burgess:2004yq,       Chialva:2005gt, Cicoli:2011yy,  Chialva:2012rq   }),
or purely stringy signatures (see, e.g. \cite{Bianchi:2006nf, Anchordoqui:2007da, Anchordoqui:2008ac, Lust:2008qc, Anchordoqui:2008di, Anchordoqui:2009ja, Lust:2009pz, Anchordoqui:2009mm,  Anchordoqui:2010zs, Feng:2010yx, Dong:2010jt, Carmi:2011dt, Hashi:2012ka, Anchordoqui:2014wha})\footnote{For recent reviews, see \cite{Lust:2013koa, Berenstein:2014wva}.}.

A series of papers \cite{Lust:2008qc,Anchordoqui:2009mm,Lust:2009pz,Anchordoqui:2009ja} focused on tree-level string scattering amplitudes containing massless bosons and fermions that can be identified with the SM fields. Amplitudes containing at most two chiral fermions were shown to exhibit a universal behavior independently of the specifics of the compactification. The  poles of these scattering amplitudes correspond to the exchange of Regge excitations of the SM gauge bosons, whose mass scales like the string mass. Due to the universal behavior of these amplitudes, which in turn gives them a predictive power, LHC is able to constrain the string scale to be above $4.5$ TeV
\cite{Chatrchyan:2011ns, ATLAS:2012pu,  Chatrchyan:2013qha}.

\begin{figure}[h]
\begin{center}
\epsfig{file=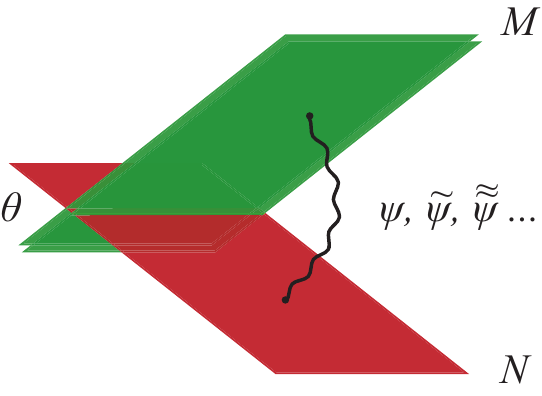,width=70mm}
\caption{Towers of states whose masses are multiples of the intersection angle live at D-brane  intersections.
}\label{fig. Intersecting Branes}
\end{center}
\end{figure}

\begin{figure}[h]
\begin{center}
\epsfig{file=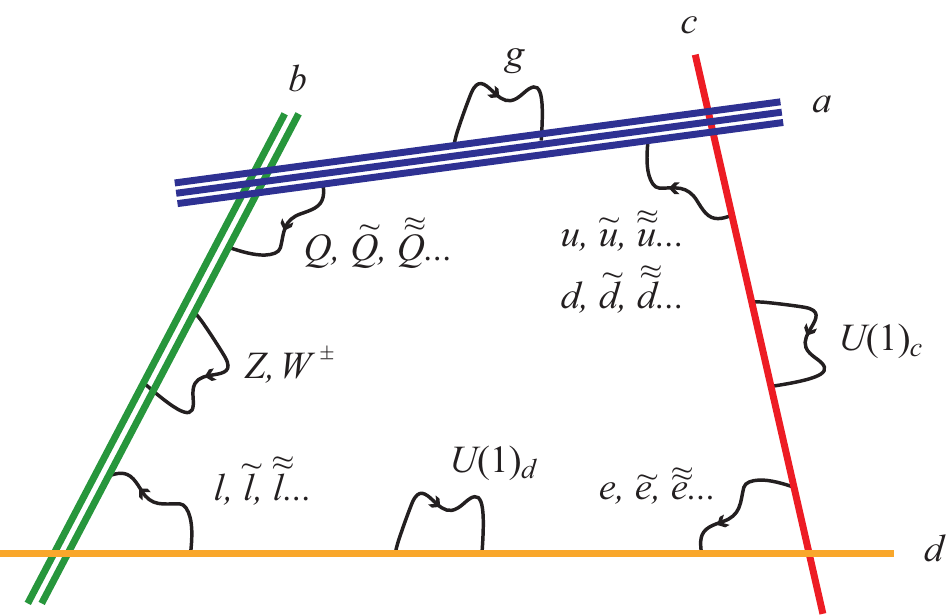,width=120mm}
\caption{Local D-brane realization of the Standard Model. Each field is accompanied by a whole tower of massive excitations.  
}\label{fig. madrid}
\end{center}
\end{figure}

In this paper, our attention goes towards a different direction. At the intersection of two D-brane stacks there exists a whole tower of stringy excitations with the same quantum numbers \cite{Anastasopoulos:2011hj, Anastasopoulos:2011gn, Anastasopoulos:2014lpa, Anastasopoulos:2015dqa}. The lowest state can be massless (for a specific value of the intersecting angles) and the rest are massive with masses proportional to the angle of the intersection in which they live (see fig. \ref{fig. Intersecting Branes}). Therefore if we call $\psi$, $\tilde \psi$, $\tilde {\tilde\psi}$ etc the states living at the intersection, we have
\bea
M^2_{\psi} = 0 ~,~~~ \alpha' M^2_{\tilde\psi} =  \theta/\pi ~,~~~ \alpha' M^2_{\tilde{\tilde\psi}} =  2 \theta/\pi ~\dots
\eea
where $\theta$ denotes the intersection angle between these two D-brane stacks. For small intersection angle those light stringy states can be significantly lighter than the first Regge excitations of the gauge bosons and are expected to be observed prior to the latter. In figure \ref{fig. madrid} we depict a D-brane SM realization\footnote{For original work on local D-brane configurations, see \cite{Antoniadis:2000ena,Aldazabal:2000sa}. For a systematic analysis of local D-brane configurations, see \cite{Gmeiner:2005vz,Anastasopoulos:2006da,Cvetic:2009yh,Cvetic:2010mm}.}. While the SM gauge bosons live on the world volume of the D-brane stacks the SM fermions are localized at the intersections of different D-brane stacks. At each intersection there exist a tower of massive stringy excitations that can be significantly lighter than the string scale for some regions in the parameter space.  

Such towers of states are different from standard string resonances or Kaluza-Klein excitations, for two main reasons. 
\begin{itemize} 
\item First of all, for standard Regge resonances the mass square grows linearly with the level and is the same for each ground-state. However, in this D-brane setup one has for each massless fermion a separate tower of stringy excitations with a different mass spacing. That gives a larger variety for the masses of the excited SM fields. 
\item Second main difference is that Kaluza-Klein scenarios forbid decays which violate momentum conservation in the compact dimension. These decays, even though with a small rate (depending on several moduli), are allowed in intersecting D-brane setups giving a lot of space for interesting phenomenology.
\end{itemize}

Aim of the present investigation is to carefully analyse the spectrum of lowest lying massive states in the case of supersymmetric intersections and provide the vertex operators for the first few excitations. We will also discuss some simple interactions and offer an alternative interpretation of the putative LHC signal at 750 GeV as a light massive open string state living at a D-brane intersection, a replica of the SM Higgs. In future works we are planning to use the toolkit developed here in order to study the peculiar phenomenology of these states and give further support to our present preliminary comments on the diphoton excess.

The paper is organized as follows. After setting up the stage in Section {2}, we compute the character-valued partition function in Section {3} and identify the physical states together with their helicity. Then, in Section {4}, we determine the relevant vertex operators and impose the conditions for BRST invariance and in Section {5} we reconstruct the super-multiplet structure. Finally, in Section {6} we compute some simple interactions, such as the decay rate of a massive scalar or vector into two massless fermions, and in Section {7} we present our conclusions and outlook, including comments on our alternative interpretation of the 750 GeV diphoton excess at LHC in terms of a massive but light Higgs replica, that couples to two photons thanks to massive string exchange.
Several appendices contain useful formulae for the twist-fields, their OPE's and state-operator correspondence.

\section{Setup}

For definiteness, our computations will be carried out for Type IIA string theory with intersecting D6-branes and $\Omega$6-planes, whose world-volume spans the four non-compact directions and a 3-cycle in some internal (Calabi-Yau) space. However, our analysis can be easily converted to Type IIB framework by T-duality. In order to ensure calculability we assume that the stacks of D6-branes supporting the Standard Model intersect with each other in a region of the internal space that looks like a factorizable torus $T^6=T^2_1\times T^2_3\times T^2_3$ \footnote{Semi-realistic MSSM constructions on factorizable orbifolds can be found in \cite{Blumenhagen:2000wh,Angelantonj:2000hi,Aldazabal:2000cn, Aldazabal:2000dg, Forste:2000hx, Ibanez:2001nd, Cvetic:2001tj, Cvetic:2001nr, 
 Honecker:2003vq, Cvetic:2004nk, Honecker:2004np, Honecker:2004kb, Blumenhagen:2004xx, Blumenhagen:2005tn, Gmeiner:2005vz, Bailin:2006zf, 
 Cvetic:2006by, Chen:2007px, Bailin:2007va,Gmeiner:2007zz, Bailin:2008xx, Gmeiner:2008xq, Honecker:2012qr, Honecker:2013kda}.}.

In terms of three complex coordinates, with $I=1,2,3$, we have
\bea
Z^{I}=X^{2I+2}+iX^{2I+3} ~,~~~ Z^*_{I}=X^{2I+2}-iX^{2I+3} 
\eea
In this framework, we consider two stacks of D6-branes which intersect at angles ${\theta}_I = \pi a_I$ in the three tori, and without loss of generality we assume that
\bea
1> {a}_1, {a}_2 , {a}_3 >0  ~.
\label{angleassumtion}\eea 
\begin{figure}[t]
\begin{center}
\epsfig{file=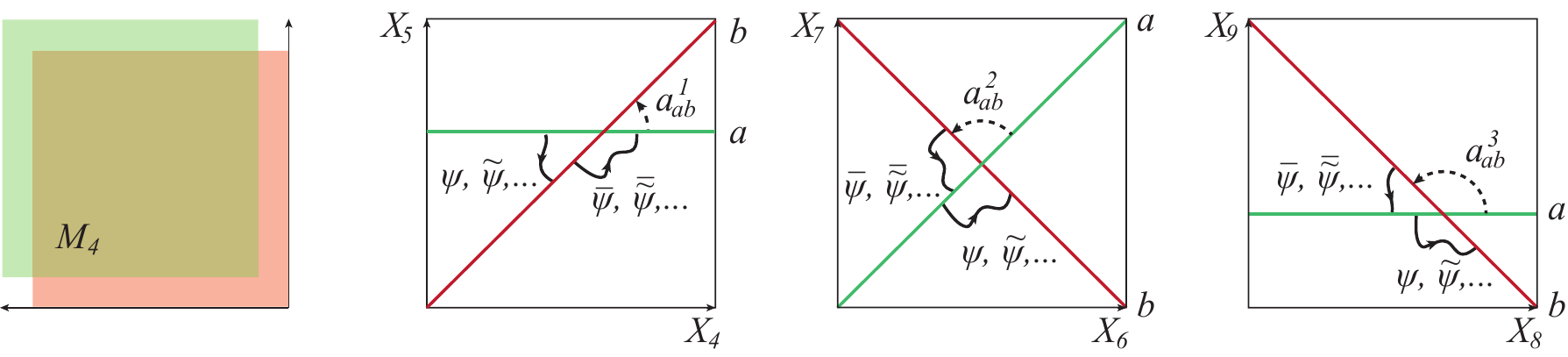,width=155mm}
\caption{At D-brane  intersections live towers of states and their masses are multiples of the intersection angle.}\label{fig. Two Intersecting Branes}
\end{center}
\end{figure}
Supersymmetry follows if we also impose
\bea
{a}_1 +{a}_2 -{a}_3  = 0 ~. \label{SUSYcondition}
\eea

In order to identify not only the massless states but also the lowest lying massive states living at the intersection between two stacks of intersecting branes, we will study the 1-loop partition function of strings being stretched between the two branes.

\section{Character valued partition function}

The 1-loop partition function of open strings living at the intersection of two stacks of branes is encoded in the vacuum amplitude on the annulus with one boundary mapped onto one stack of ${N}$ branes and the other onto another stack of ${M}$ branes. All the states transform according to the bi-fundamental representation $({\bf N}, {\bf \overline{M}}) $ of the gauge group $U({N}){\times} U({M})$. For simplicity we will not consider here un-oriented strings that can appear in (anti)symmetric representations of unitary as well as orthogonal or symplectic groups. These would require the inclusion of $\Omega$-planes and as a result of the corresponding M\"obius-strip amplitudes.  
In order to keep track of the helicity of each state, we consider the character-valued one-loop partition function or helicity super-trace, that reads
\bea
{\cal Z} = {\rm STr} (z^h q^{\alpha' {m}^2})= |{\cal I}| i\sin\pi v \sum_a c_a^{^{GSO}} {\vartheta_a(v) \prod_I \vartheta_a(u_I) \over \vartheta_1(v) \prod_I \vartheta_1(u_I)}
\eea
where $q=\exp 2\pi i \tau$, $z=\exp 2\pi i v$, $h$ denotes the helicity and $|{\cal I}|$ is the degeneracy of the ground-state,
{\it i.~e.} number of intersections or of Landau levels (in the T-dual case of magnetised branes). We also set $u_I = {a}_I  \tau$ where $\tau = iT/2$ for the annulus\footnote{For the M\"obius-strip one would have $\tau = iT/2 + 1/2$}. The multiplicity $ N\overline{M}$ is understood. The GSO projection is implemented by the sum over the spin structures with $c_3^{^{GSO}} = - c_4^{^{GSO}} = - c_2^{^{GSO}} = 1$ and $c_1^{^{GSO}} = \pm 1$.
The sign of ${\cal I}$ in combination with the sign of $c_1^{^{GSO}}$ determines the chirality of the massless fermion always present in the ground-state. 
The $odd$ spin structure simply contributes
\bea
{\cal Z}_{\rm odd} = \pm |{\cal I}| i\sin\pi v {\vartheta_1(v) \prod_I \vartheta_1(u_I) \over \vartheta_1(v) \prod_I \vartheta_1(u_I)} = \pm |{\cal I}| i\sin\pi v  = \pm {1\over 2} |{\cal I}| (z^{+1/2} - z^{-1/2})
\eea
Additional bosonic and fermionic zero-modes may appear depending on the values of  the ``angles'' ${\theta}_I = \pi {a}_I $.  
After using Riemann identity, the contribution of the $even$ spin structures becomes
\bea
{\cal Z}_{\rm even} = {i \sin\pi v |{\cal I}| \over  \vartheta_1(v) \prod_I \vartheta_1(u_I)}
 \sum_{\epsilon=\pm 1} \vartheta_1\left({1\over 2} (v+\epsilon \sum_I u_I)\right) \prod_K \vartheta_1\left({1\over 2} (v-\epsilon \sum_I u_I) + \epsilon u_K \right)\nn\\
\label{Zeven} \eea
Assuming $1\gg {a}_I  \ge 0$ as well as the ``triangular inequality" $\sum_I {a}_I  \ge 2 {a}_K$, using Riemann identity for Jacobi $\vartheta$ functions, provided in appendix \ref{Theta functions and Riemann identity}, and neglecting all oscillators with $\nu=n$, $n+\sum_I\epsilon_I{a}_I $ with $n\ge 1$, we get only states with spin $s\le 1$
\bea
{\cal Z}_{\rm even} &=& -{|{\cal I}| \over 2 \prod_I (1 - q^{{a}_I })} 
\Big\{(z^{1/2} - z^{1}q^{\frac{1}{2}({a}_1 +{a}_2 +{a}_3 )}) 
(1 - z^{-1/2}q^{\frac{1}{2}(-{a}_1 +{a}_2 +{a}_3 )})\label{Zeven second form}\\
&&
~~~~~~~~~~~~~~~~~~~~~ (1 - z^{-1/2}q^{\frac{1}{2}({a}_1 -{a}_2 +{a}_3 )})
(1 - z^{-1/2}q^{\frac{1}{2}({a}_1 +{a}_2 -{a}_3 )})  +(z\to z^{-1})\Big\}\nn\\
&=&  { |{\cal I}| \over 2 \prod_I (1 - q^{{a}_I })} \Big\{
2\Big(q^{\frac{1}{2}({a}_1 +{a}_2 -{a}_3 )} + q^{\frac{1}{2}({a}_1 -{a}_2 +{a}_3 )} +q^{\frac{1}{2}({a}_1 +{a}_2 -{a}_3 )} \nn\\
&& ~~~~~~~~~~~~~~~~~~~~~~~~~~~~ +q^{\frac{1}{2}(3{a}_1 +{a}_2 +{a}_3 )} + q^{\frac{1}{2}({a}_1 +3{a}_2 +{a}_3 )} + q^{\frac{1}{2}({a}_1 +{a}_2 +3{a}_3 )}\Big) \nn\\
&& ~~~~~~~~~~~~~~~~~~~ -(1+q^{{a}_1 })(1+q^{{a}_2 })(1+q^{{a}_3 })\Big(z^{+1/2} + z^{-1/2}\Big) \nn\\
&& ~~~~~~~~~~~~~~~~~~~~~~~~~~~~~~~~~~~~~~~~~~~~~~~~~~ + 2 q^{\frac{{a}_1 +{a}_2 +{a}_3 }{2}} \Big(z^{+1} + z^{-1}\Big) + \ldots \Big\} ~~~~~\nn
\eea
Notice that the above formula is valid for any $a_I>0$.

\subsubsection*{Supersymmetric case ${a}_1 +{a}_2 -{a}_3 =0$}

For definiteness, we henceforth enforce the supersymmetric condition\footnote{Any combination ${a}_1\pm{a}_2\pm{a}_3 =0$ would preserve some supersymmetry.} ${a}_1 +{a}_2 -{a}_3 =0$. Using the expansion of the geometric series
%
%
%
$
1/({1-q^{a_I}}) = 1 + q^{a_I} + q^{2a_I} + q^{3a_I} + \dots
$
for small angles $a_I \ll 1$ and assuming $a_3 = {a}_1 +{a}_2 > {a}_2 \ge {a}_1$, the total partition function becomes 
\begin{equation}
\begin{split}
{\cal Z}_\textup{all}&=(1-z^{\mp 1/2})q^0+(2-z^{-1/2}-z^{1/2})(q^{a_1}+q^{a_2})+\\
&+(2-z^{-1/2}-z^{1/2})(q^{2 a_1}+q^{2 a_2})+(z^{-1}-3 z^{-1/2}+4-3 z^{1/2}+z^{1}) q^{a_1+a_2}+\\
&+(2-z^{-1/2}-z^{1/2})(q^{3 a_1}+q^{3 a_2})+(z^{-1}-4 z^{-1/2}+6-4 z^{1/2}+z^1)(q^{2 a_1+a_2}+q^{a_1+2 a_2})+\\
&+(2-z^{-1/2}-z^{1/2})(q^{4 a_1}+q^{4 a_2})+(z^{-1}-4 z^{-1/2}+6-4 z^{1/2}+z^1)(q^{3 a_1+a_2}+q^{a_1+3 a_2})+\\
&+(2 z^{-1}-7 z^{-1/2}+10-7 z^{1/2}+2 z^1)q^{2 a_1+2 a_2}+\dots
\end{split}
\end{equation}
Therefore, the states living at the intersection of the two branes are organized in multiplets with different masses:
\begin{itemize}
\item Massless sector (from $q^0$)
\bea
{\cal Z}^{\rm massless}_{\rm chiral} = |{\cal I}| (1-z^{\pm 1/2})
\eea
produces a massless (anti-)chiral multiplet.

\item Massive states with $\a'  {m}^2=k{a}_{1,2}$ (from $q^{k{a}_{1,2}}$ with $k=1,2,\ldots$)
\bea
{\cal Z}^{\a'  {m}^2=k {a}_{1,2}}
& =& |{\cal I}| 
\Big(2 - z^{-1/2} -z^{1/2}\Big) q^{k {a}_{1,2}}
\eea
form massive chiral multiplets (one chiral and one anti-chiral, as we will see).

\item Massive states with $\a'  {m}^2 = k a_1 + a_2 = a_3 + (k{-}1)a_1$ (from $q^{k a_1 + a_2}$ with $k=1,2,\ldots$)
\bea
{\cal Z}^{\a'  {m}^2= a_3 + (k{-}1)a_1}
& =& |{\cal I}| 
\Big\{\Big(2 - z^{-1/2} -z^{1/2}\Big)+\Big(2 - 2 z^{-1/2} -2 z^{1/2} + z^{-1} + z^{1}\Big)\Big\} q^{k a_1 + a_2}\nn\\
\eea
form a massive chiral and a massive vector multiplet. The same applies to states with 
$\a'  {m}^2={a}_1+k{a}_2 = a_3 + (k{-}1)a_2$.

\item Massive states with $\a'  {m}^2=2{a}_3 = 2{a}_1 + 2{a}_2$ (from $q^{2{a}_1 + 2{a}_2}$)
\bea
{\cal Z}^{\a'  {m}^2=2{a}_3}
& =& |{\cal I}| 
\Big\{3\Big(2 - z^{-1/2} -z^{1/2}\Big)+2 \Big(2 - 2 z^{-1/2} -2 z^{1/2} + z^{-1} + z^{1}\Big)\Big\} q^{2{a}_3}\nn\\
\eea
form three massive chiral and two massive vector multiplets.

\end{itemize}

The precise identification of the lowest lying physical states allows us to determine their vertex operators: a task we will accomplish in the next section.

\section{States and vertex operators}

Let us write down the vertex operators (VO's) for the massless and lowest-lying massive states with $\a'  {m}^2={a}_I$ with $I=1,2,3$, starting from the massless ones.

\subsection{Gauge sector}

Recall that the Left-Handed (LH) and Right-Handed (RH) gaugino vertex operators in the canonical $-1/2$ super-ghost picture read
\bea
V_\lambda = \lambda^\alpha(k) S_\alpha \prod_{I=1}^3 e^{i\varphi_I/2} e^{-\varphi/2} e^{ipX}
\quad , \quad 
V_{\tilde\lambda} = \tilde\lambda^{\dot\alpha}(k) C_{\dot\alpha} \prod_{I=1}^3 e^{-i\varphi_I/2} e^{-\varphi/2} e^{ipX}
\eea
where $S_\alpha$ and $C_{\dot\alpha}$ denote $SO(1,3)$ spin fields of opposite chirality. They transform in the adjoint representation ${\bf N}\overline{\bf N}{+}{\bf M}\overline{\bf M}$ of the gauge group 
and determine the unbroken SUSY charges 
\bea
Q^{(-1/2)}_\alpha = \oint \frac{dz}{2 \pi i} S_\alpha \prod_{I=1}^3 e^{i\varphi_I/2} e^{-\varphi/2} 
\quad , \quad 
\tilde{Q}^{(-1/2)}_{\dot\alpha} = \oint \frac{dz}{2 \pi i} C_{\dot\alpha} \prod_{I=1}^3 e^{-i\varphi_I/2} e^{-\varphi/2} 
\eea
that should act ``locally" on all physical vertex operators (only integer powers in the OPE). As a result the three `internal' $U(1)$ R-charges $r_I$ are quantised. For instance $\sum_I r_I = -1/2$ for massless LH matter fermions in the canonical super ghost picture, as we will see momentarily. For later use, let us recall the form of the SUSY charges in the +1/2 super-ghost picture
\bea
&& Q^{(+1/2)}_\alpha =  \oint \frac{dz}{2 \pi i} e^{+\varphi/2}\Big( S_\alpha \partial Z^I \psi^*_I + \partial X_\mu \sigma^\mu_{\alpha\dot\alpha} C^{\dot\alpha}\Big) \prod_{I=1}^3 e^{i\varphi_I/2}    \\
&& \tilde{Q}^{(+1/2)}_{\dot\alpha} = \oint \frac{dz}{2 \pi i} e^{+\varphi/2} \Big( C_{\dot\alpha}\partial Z^*_I \psi^I + \partial X_\mu S^\alpha\sigma^\mu_{\alpha\dot\alpha} \Big) \prod_{I=1}^3 e^{-i\varphi_I/2}  
\eea

\subsection{Massless sector}

For a choice of GSO and sign of ${\cal I}$, the massless chiral multiplets, living at the $|{\cal I}|$  intersections and transforming in the $(\bf{N},\overline{\bf{M}})$ representation, consist in one scalar and a LH fermion each, whose internal states read 
\bea
\phi(k)  ~:~~ \psi^3_{-\frac{1}{2}+a_3} |a_1,a_2, a_3 \rangle_{NS} ~,~~~~~~~~~
\chi^\alpha(k) ~:~~~ |a_1, a_2, a_3 \rangle_R
\eea
where\footnote{Our notation here differs from the one in \cite{Anastasopoulos:2011hj} since $ |a_1,a_2, a_3 \rangle^{here}_{NS} \leftrightarrow |a_1,a_2, - a_3 \rangle^{there}_{NS}$. One also has $\sigma_{a}^\dagger = \sigma_{1-a}$.} $ |a_1,a_2, a_3 \rangle_{NS} = \lim_{z\rightarrow 0} \sigma_{{a}_1 } \sigma_{{a}_2 } \sigma^\dagger_{{a}_3 } (z) |0 \rangle_{NS} $ 
and vertex operators (VO's henceforth) \footnote{For a detailed discussion on vertex operators of massless states for arbitrary intersection angles,
see \cite{Cvetic:2006iz,Bertolini:2005qh}, for a generalization to massive states, see \cite{Anastasopoulos:2011hj}, and for a discussion on instantonic states
at the intersection of D-instanton and D-brane at arbitrary angles, see \cite{Cvetic:2009mt}.}
\bea
&&V^0_\phi = \phi(k) \sigma_{{a}_1 } \sigma_{{a}_2 } \sigma^\dagger_{{a}_3 } 
e^{i[{a}_1 \varphi_1 + {a}_2 \varphi_2 + (1-{a}_3 ) \varphi_3]} e^{-\varphi} e^{ikX}\\
&&V^0_\chi = \chi^\alpha(k) S_\alpha \sigma_{{a}_1 } \sigma_{{a}_2 } \sigma^\dagger_{{a}_3 } 
e^{i[({a}_1 - {1\over 2})\varphi_1 + ({a}_2 - {1\over 2})\varphi_2 + ({1\over 2} - {a}_3  ) \varphi_3]} e^{-\varphi/2} e^{ikX}
\eea
The GSO projection determines the chirality of the four-dimensional spinor in terms of the $U(1)$ world-sheet charge of the respective vertex operator. The latter is given by $\sum ^3_{I=1} r_I$, where the $r_I$ are the coefficients of $\varphi_I$, that bosonize the internal world-sheet fermions, {\it i.~e.} $\psi^I = \exp(i\varphi_I)$, $\psi^*_I = \exp(-i\varphi_I)$. If the sum adds up to $-1/2~({\rm mod}~2)$ the spinor is chiral (LH),
while if the sum is $+1/2~({\rm mod}~2)$ the spinor is anti-chiral (RH). Similarly complex bosons in chiral multiplets, pairing with LH fermions, carry $\sum ^3_{I=1} r_I = +1$, while those in anti-chiral multiplets, pairing with RH fermions, carry $\sum ^3_{I=1} r_I = -1$. In a not-so-lose sense this charge can be identified with the R-charge of the ${\cal N} =1$ SUSY algebra in that -- in the canonical super- ghost picture -- 
the super-charges and the gaugini carry R-charges $\pm 3/2$. Changing picture vertex operators and super-charges have no definite R-charge, which is not an exact continuous symmetry in fact.

The BRST charge is (focusing on the matter part)
\begin{align}
Q_{BRST}&= \oint \frac{dz}{2 \pi i}\bigg\{ e^{\phi} \frac{ \eta}{\sqrt{2 \alpha'}} \bigg( i\partial X^{\mu} \, \psi_{\mu} + \sum^3_{I=1} \partial Z^{I } \, \psi^*_I + \sum^3_{I=1} \partial Z^*_{I } \, \psi^I\bigg) \label{eq BRST charge}\\
&~~~~+ \frac{c}{\a'} \bigg[i\partial X^\mu i \partial X_\mu - \frac{\a'}{2} \psi^\mu \partial \psi_\mu - \sum^3_{I=1} \Big(\partial Z^I \partial Z^*_I + \frac{\a'}{2} \psi^I \partial \psi^*_I + \frac{\a'}{2} \psi^*_I \partial \psi^I \Big)\bigg] \bigg\} + ...\nn
\end{align}
and the physical state condition is
\bea
[Q_{BRST}, V] = 0
\label{BRST condition}\eea
and using the OPE's we get a $simple$ and a $double$ pole. 
\begin{itemize} 
\item The simple pole identically vanishes for the scalar. For the fermions, the vanishing of the simple pole requires the Dirac equation of motion. In this massless case it is just
\bea
k^\mu \bar\sigma_\mu^{\dot\alpha\alpha} \chi_\alpha(k) = 0
\eea

\item Vanishing of the double pole always yields the standard mass-shell condition 
\bea 
k^2=0 
\eea
for any massless bosonic or fermionic VO. For massive ones one gets 
\bea 
p^2+m^2 = 0
\eea
Henceforth we will use $k$ to denote light-like momenta and $p$ to denote time-like ones. Dependence on the momenta of the `polarisations' or `wave-functions' of the vertex operators will be explicitly shown as $\ell (k)$ or $H(p)$ for convenience.
\end{itemize} 

\subsection{Massive multiplets}

Here we study separately the $\alpha' {m}^2 = {a}_{1,2}$ and the $\alpha' {m}^2 = {a}_3$ cases. LH fermions in chiral multiplets in the $({\bf N}, {\bf \bar{M}})$ representation will be denoted by $\chi^I_f(p)$, their scalar partners by $\phi^I_f(p)$ with $I=1,2,3$ referring to their mass $\alpha' {m}^2 = {a}_{I}$ and $f=1, ...n_I$ their multiplicity. LH fermions and scalars in chiral multiplets in the $({\bf M}, {\bf \bar{N}})$ will be denoted by $\tilde\chi^{I}_f(p)$, their scalar partners by $\tilde\phi^I_f(p)$. States in anti-chiral multiplets will be distinguished by a dagger.

\subsubsection{Massive multiplets with $\alpha' {m}^2 = {a}_{1,2}$}

Two massive scalar multiplets (2 scalars and one Dirac fermion) each with masses $\alpha' {m}^2 = {a}_{1, 2}$

\begin{itemize}

\item Scalars with mass $\alpha' {m}^2 = {a}_1$.

We have two state with this mass: 
\bea
\tilde\phi_1^{1\dagger} ~:~~~ \psi^{{2}^*}_{-\frac{1}{2}+a_2} |a_1,a_2,a_3 \rangle_{NS}
~,~~~~~~
 \phi^1_2 ~:~~~  {\alpha}^1_{-a_1} \psi^3_{-\frac{1}{2}+a_3} |a_1,a_2, a_3 \rangle_{NS}
\eea
their VO's are\footnote{The boson ${\tilde\phi_1^{1\dagger}}$ has R-charge $R=-1={a}_1 +{a}_2 -1-{a}_3 $ and it may be thought of as obtained by acting with $\psi^*_2$ on $e^{i[{a}_1 \varphi_1 + {a}_2 \varphi_2 - {a}_3 \varphi_3]}$.  Similarly the RH fermion (see below) has R-charge $R=+1/2=({a}_1 -1/2) +({a}_2 +1/2)+(1/2-{a}_3 )$ and it may be thought as obtained by acting with $\psi^*_2$ on $\Sigma_{+3/2} = e^{i[({a}_1 +1/2)\varphi_1 + ({a}_2 +1/2)\varphi_2 - ({a}_3 -1/2)                                                                                                                             \varphi_3]}$.} 
\bea 
&&V^{{a}_1}_{\tilde\phi_1^{1\dagger}} = \tilde\phi_1^{1\dagger}{(p)} \sigma_{{a}_1 } \sigma_{{a}_2 } \sigma_{{a}_3 }^\dagger 
e^{i[{a}_1 \varphi_1 + ({a}_2 -1)\varphi_2 - {a}_3  \varphi_3]} e^{-\varphi} e^{ipX}\\
&&V^{{a}_1}_{\phi^1_2} =\frac{1}{\sqrt{{a}_1 }} \phi^1_{2}{(p)} \tau_{{a}_1 } \sigma_{{a}_2 } \sigma_{{a}_3 }^\dagger 
e^{i[{a}_1 \varphi_1 + {a}_2 \varphi_2 +(1- {a}_3 ) \varphi_3]} e^{-\varphi} e^{ipX}
\eea
The appearance of a normalisation factor $\frac{1}{\sqrt{{a}_1 }}$ in front of $\tau_{a_1}$ here and henceforth is to simplify OPE formulae. Checking with BRST charge we get zero simple pole and the double pole gives 
\bea
\alpha' p^2|_{\tilde\phi^{1\dagger}_1} = {a}_1  ~,~~~~~ \alpha' p^2|_{\phi^1_2} = \frac{3{a}_1  + {a}_2  - {a}_3 }{2} = {a}_1  
\eea
Therefore, these two states can be combined to give one massive complex scalar.
They are both complex since they belong to the representation $({\bf N, \bar{M}})$ of the Chan-Paton group. The first one $\tilde\phi_1^{1\dagger}$ has R-charge $-1$ (to be precise $[0-10]$. Here and henceforth we expose the three R-charges as $[2{r}_1, 2{r}_2, 2{r}_3]:= [a_1+ {r}_1, a_2+ {{r}_2}, - a_3+ {{r}_3}]$ so that $R= {r}_1+{r}_2+{r}_3$ since $a_1+a_2-a_3=0$.) so it is better thought of as the complex conjugate of the scalar $\tilde\phi_1^{1}$ with R-charge $+1$ (to be precise [001]) in the $({\bf \bar{N}, M})$ arising from strings with the opposite orientation. The gauge invariant super-potential term is $W = m_1 \tilde\phi_1^{1} \phi^1_{2} $  where $m_1 = \sqrt{a_1/\alpha'} = \langle M_1\rangle$ may be thought of as the VEV of a chiral closed string singlet $M_1$ with R-charges [100].

\item Fermions with mass $\alpha' {m}^2 = {a}_1$.

We have again two states with this mass: 
\bea
\tilde \chi_1^1 ~:~~~ 
\psi^1_{-a_1} |a_1,a_2,a_3\rangle_R
~,~~~~~~
 \chi^1_2 ~:~~~  
{\alpha}^1_{-a_1} |a_1,a_2,a_3\rangle_R
\eea

The $\tilde \chi_1^1$ and the string with opposite orientation $\psi^{{1}^*}_{-a_1} |1-a_1,1-a_2,1- a_3\rangle_R$ have VO's
\bea
&& V^{{a}_1}_{\tilde\chi_1^1} = (\tilde\chi^1_1)^{\dot\alpha}{(p)} C_{\dot\alpha} \sigma_{{a}_1 } \sigma_{{a}_2 } \sigma_{{a}_3 }^\dagger 
e^{i[({a}_1 +{1\over 2}) \varphi_1 + ({a}_2  - {1\over 2})\varphi_2 - ({a}_3 -{1\over 2}) \varphi_3]} e^{-\varphi} e^{ipX}\\
&& V^{{a}_1}_{\chi_1^1} = (\chi^1_1)^{\alpha}{(p)} S_{\alpha} \sigma^\dagger_{{a}_1 } \sigma_{{a}_2 }^\dagger \sigma_{{a}_3 } 
e^{-i[({a}_1 +{1\over 2}) \varphi_1 + ({a}_2  - {1\over 2})\varphi_2 - ({a}_3 -{1\over 2}) \varphi_3]} e^{-\varphi} e^{ipX} ~~~~~
\eea
The $\chi_2^1$ and the string with opposite orientation ${\alpha}^{{1}^*}_{-a_1} |1-a_1,1-a_2,1-a_3\rangle_R$ have VO's: 
\bea
&&V^{{a}_1}_{\chi_2^1} = \frac{1}{\sqrt{{a}_1 }} (\chi_2^1)^{\alpha}{(p)} S_{\alpha} \tau_{{a}_1 } \sigma_{{a}_2 } \sigma_{{a}_3 }^\dagger 
e^{i[({a}_1 -{1\over 2}) \varphi_1 + ({a}_2  - {1\over 2})\varphi_2 - ({a}_3 -{1\over 2}) \varphi_3]} e^{-\varphi} e^{ipX}\\
&&V^{{a}_1}_{\tilde \chi_2^1} =  \frac{1}{\sqrt{{a}_1 }} (\tilde \chi_2^1)^{\dot \alpha}{(p)} C_{\dot\alpha} \tau^\dagger_{{a}_1 } \sigma_{{a}_2 }^\dagger \sigma_{{a}_3 } 
e^{-i[({a}_1 -{1\over 2}) \varphi_1 + ({a}_2  - {1\over 2})\varphi_2 - ({a}_3 -{1\over 2}) \varphi_3]} e^{-\varphi} e^{ipX} ~~~~~
\eea

The BRST condition \eqref{BRST condition} apart from the usual double pole which vanishes for $a' p^2 = {a}_1$ gives also a simple pole which vanishes if we combine the two VO's like
\bea
V= V^{{a}_1}_{\tilde\chi_1^1} + V^{{a}_1}_{\chi_2^1}     ~~~~&\sim&~~~~  ( (\tilde \chi_1^1)_{\dot\alpha}{(p)} + \frac{{p_\mu} (\chi_2^1)^{\alpha}{(p)}  \sigma^\mu_{\alpha\dot\alpha}}{\sqrt{{a}_1 }})~C^{\dot\alpha} ~~~~~~ \nn\\
~~~~~~~~~~~~~~~~~~~~~&\to&~~~~ {\sqrt{{a}_1 }}   (\tilde \chi_1^1)_{\dot\alpha}{(p)} + {{p_\mu} (\chi_2^1)^{\alpha}{(p)}  \sigma^\mu_{\alpha\dot\alpha}}= 0 ~~~~~~
\eea
which is the Dirac equation for a fermion\footnote{Notice that the normalization of the combined VO comes to cancel the coefficient in the OPE.} $\Big((\chi_2^1)^{\alpha}, (\tilde \chi_1^1)_{\dot\alpha}\Big)$.

\end{itemize}
%
%
%
Scalars and fermions with mass $\alpha' {m}^2 = {a}_2$ can be obtained from the above by the exchange $1\leftrightarrow 2$.

\subsubsection{Massive multiplets with $\alpha' {m}^2 = {a}_3$}

We also have several states with mass $\alpha' {m}^2 = {a}_3$.
\begin{itemize}

\item Scalars with mass $\alpha' {m}^2 = {a}_3$.

We find four scalars with this mass
\bea
&& \phi_1^3 ~:~~~ {\alpha}^{{3}^*}_{-a_3} \psi^3_{-\frac{1}{2}+a_3} |a_1,a_2, a_3 \rangle_{NS}~,~~~~~~
 \phi^3_2 ~:~~~  {\alpha}^1_{-a_1} {\alpha}^2_{-a_2} \psi^3_{-\frac{1}{2}+a_3} |a_1,a_2, a_3 \rangle_{NS} ~~~~~~~~~~\\
&& \tilde\phi_3^{3\dagger} ~:~~~  {\alpha}^1_{-a_1} \psi^{{1}^*}_{-\frac{1}{2}+a_1} |a_1,a_2, a_3 \rangle_{NS}~,~~~~~~
 \tilde\phi^{3\dagger}_4 ~:~~~   {\alpha}^2_{-a_2} \psi^{{2}^*}_{-\frac{1}{2}+a_2} |a_1,a_2, a_3 \rangle_{NS} 
\eea
Their VO's are
\bea 
&& V^{{a}_3}_{\phi^3_1} =\frac{1}{\sqrt{{a}_3 }} \phi^3_{1}{(p)} \sigma_{{a}_1 } \sigma_{{a}_2 } \tau_{{a}_3 }^\dagger  
e^{i[{a}_1 \varphi_1 + {a}_2 \varphi_2 +(1- {a}_3 ) \varphi_3]} e^{-\varphi} e^{ipX}\\
&& V^{{a}_3}_{\phi^3_2} = \frac{1}{\sqrt{{a}_1 {a}_2 }}\phi^3_{2}{(p)} \tau_{{a}_1 } \tau_{{a}_2 } \sigma_{{a}_3 }^\dagger 
e^{i[{a}_1 \varphi_1 + {a}_2 \varphi_2 +(1- {a}_3 ) \varphi_3]} e^{-\varphi} e^{ipX}\\
&& V^{{a}_3}_{ \tilde\phi_3^{3\dagger}} = \frac{1}{\sqrt{{a}_1 }}\phi^3_{3}{(p)} \tau_{{a}_1 } \sigma_{{a}_2 } \sigma_{{a}_3 }^\dagger 
e^{i[({a}_1 -1)\varphi_1 + {a}_2 \varphi_2 +{a}_3  \varphi_3]} e^{-\varphi} e^{ipX}\\
&& V^{{a}_3}_{\tilde\phi^{3\dagger}_4} = \frac{1}{\sqrt{{a}_2 }}\phi^3_{4}{(p)} \sigma_{{a}_1 } \tau_{{a}_2 } \sigma_{{a}_3 }^\dagger 
e^{i[{a}_1 \varphi_1 + ({a}_2 -1)\varphi_2 +{a}_3  \varphi_3]} e^{-\varphi} e^{ipX}
\eea
Commutation with the BRST charge produces $single$ poles
\bea
&& [Q_{BRST},V^{{a}_3}_{\phi^3_1} ] ~~\to~~ \frac{1}{z} \sqrt{{a}_3 } \sigma_{{a}_1 } \sigma_{{a}_2 } \sigma_{{a}_3 }^\dagger  \\
&& [Q_{BRST},V^{{a}_3}_{\phi^3_2} ]  ~~\to~~ \frac{1}{z} 0 \\
&& [Q_{BRST},V^{{a}_3}_{\tilde\phi^{3\dagger}_3} ]  ~~\to~~ \frac{1}{z} \sqrt{{a}_1 } \sigma_{{a}_1 } \sigma_{{a}_2 } \sigma_{{a}_3 }^\dagger  \\
&& [Q_{BRST},V^{{a}_3}_{\tilde\phi^{3\dagger}_4} ] ~~\to~~ \frac{1}{z} \sqrt{{a}_2 } \sigma_{{a}_1 } \sigma_{{a}_2 } \sigma_{{a}_3 }^\dagger 
\eea
as well as $double$ poles that vanish for the usual $\alpha ' p^2={a}_3 $ mass-shell condition. 

Therefore, the four complex scalars yield three BRST invariant combinations $\phi^3_2$, $\tilde\phi^{3\dagger}_3 + \tilde\phi^{3\dagger}_4$ and $a_3\phi^3_1 + a_1\tilde\phi^{3\dagger}_3 - a_2 \tilde\phi^{3\dagger}_4$. The orthogonal linear combination $-a_3\phi^3_1 + a_1\tilde\phi^{3\dagger}_3 - a_2 \tilde\phi^{3\dagger}_4$ is not BRST invariant, and it mixes with the longitudinal component of the massive vector $p{\cdot}W$ as we will momentarily see.

\item Fermions with mass $\alpha' {m}^2 = {a}_3$.

We find six fermions
\bea
&& \tilde\chi_1^3 ~:~~~ \psi^{{3}^*}_{-a_3} |a_1,a_2, a_3 \rangle_{R}~,~~~~~~~~~~~~
\chi^3_2 ~:~~~  {\alpha}^{{3}^*}_{-a_3} |a_1,a_2, a_3 \rangle_{R} ~~~~~~~~~~\\
&& \chi_3^3 ~:~~~  \psi^{{1}}_{-a_1} \psi^{{2}}_{-a_2} |a_1,a_2, a_3 \rangle_{R}~,~~~~~~
\tilde\chi^3_4 ~:~~~   {\alpha}^{{1}}_{-a_1} \psi^{{2}}_{-a_2} |a_1,a_2, a_3 \rangle_{R}\\ 
&& \tilde\chi_5^3 ~:~~~  \psi^{{1}}_{-a_1} {\alpha}^{{2}}_{-a_2} |a_1,a_2, a_3 \rangle_{R}~,~~~~~~
\chi^3_6 ~:~~~   {\alpha}^{{1}}_{-a_1} {\alpha}^{{2}}_{-a_2} |a_1,a_2, a_3 \rangle_{R} 
\eea
Their VO's are 
\bea 
&& V^{{a}_3}_{\tilde\chi^3_1} = (\tilde \chi_1^3)^{\dot\alpha}{(p)} C_{\dot\alpha} \sigma_{{a}_1 } \sigma_{{a}_2 } \sigma^\dagger_{{a}_3 } 
e^{i[({a}_1 - {1\over 2})\varphi_1 + ({a}_2 - {1\over 2})\varphi_2 - ({a}_3  +{1\over 2} ) \varphi_3]} e^{-\varphi/2} e^{ipX}~~~~\\
&&V^{{a}_3}_{\chi^3_2} = \frac{1}{\sqrt{{a}_3 }} (\chi_2^3)^{\alpha}{(p)} S_{\alpha} \sigma_{{a}_1 } \sigma_{{a}_2 } \tilde\tau^\dagger_{{a}_3 } 
e^{i[({a}_1 - {1\over 2})\varphi_1 + ({a}_2 - {1\over 2})\varphi_2 + ({1\over 2} - {a}_3  ) \varphi_3]} e^{-\varphi/2} e^{ipX}\\
&&V^{{a}_3}_{\chi^3_3} = (\chi_3^3)^{\alpha}{(p)} S_{\alpha} \sigma_{{a}_1 } \sigma_{{a}_2 } \sigma^\dagger_{{a}_3 } 
e^{i[({a}_1 +{1\over 2})\varphi_1 + ({a}_2  +{1\over 2})\varphi_2 + ({1\over 2} - {a}_3  ) \varphi_3]} e^{-\varphi/2} e^{ipX}\\
&&V^{{a}_3}_{\tilde\chi^3_4} = \frac{1}{\sqrt{{a}_1 }} (\tilde\chi_4^3)^{\dot\alpha}{(p)} C_{\dot\alpha} \tau_{{a}_1 } \sigma_{{a}_2 } \sigma^\dagger_{{a}_3 } 
e^{i[({a}_1 - {1\over 2})\varphi_1 + ({a}_2 +{1\over 2})\varphi_2 + ({1\over 2} - {a}_3  ) \varphi_3]} e^{-\varphi/2} e^{ipX}\\
&&V^{{a}_3}_{\tilde\chi^3_5} = \frac{1}{\sqrt{{a}_2 }} (\tilde\chi_5^3)^{\dot\alpha}{(p)} C_{\dot\alpha} \sigma_{{a}_1 } \tau_{{a}_2 } \sigma^\dagger_{{a}_3 } 
e^{i[({a}_1 +{1\over 2})\varphi_1 + ({a}_2 - {1\over 2})\varphi_2 + ({1\over 2} - {a}_3  ) \varphi_3]} e^{-\varphi/2} e^{ipX}\\
&&V^{{a}_3}_{\chi^3_6} = \frac{1}{\sqrt{{a}_1 {a}_2 }} (\chi_6^3)^{\alpha}{(p)} S_{\alpha} \tau_{{a}_1 } \tau_{{a}_2 } \sigma^\dagger_{{a}_3 } 
e^{i[({a}_1 - {1\over 2})\varphi_1 + ({a}_2 - {1\over 2})\varphi_2 + ({1\over 2} - {a}_3  ) \varphi_3]} e^{-\varphi/2} e^{ipX} ~~~~~~~~
\eea

Commutation with the BRST charge produces $single$ poles and $double$ poles. Combining all VO's together, collecting coefficients of different terms and imposing the physical state condition we get 
\bea
&& {\alpha' {p_\mu} (\tilde \chi_1^3)_{\dot\alpha}{(p)} \bar\sigma_\mu^{\dot\alpha\alpha}} - {\sqrt{{a}_3 }}   (\chi_2^3)^{\alpha}{(p)}= 0 ~~~\\
&& {\frac{\alpha' {p_\mu}}{\sqrt{{a}_3 }} (\chi_2^3)^{\alpha}{(p)}  \sigma^\mu_{\alpha\dot\alpha}} + (\tilde \chi_1^3)_{\dot\alpha}{(p)} = 0 ~~~\\
&&  {\alpha' {p_\mu} (\chi_3^3)^{\alpha}{(p)}  \sigma^\mu_{\alpha\dot\alpha}} + (\tilde \chi_4^3)_{\dot\alpha}{(p)} + (\tilde \chi_5^3)_{\dot\alpha}{(p)} = 0 ~~~\\
&& {\alpha' {p_\mu} (\tilde \chi_4^3)_{\dot\alpha}{(p)} \bar\sigma_\mu^{\dot\alpha\alpha}} - a_1 (\chi_3^3)^{\alpha}{(p)} + (\chi_6^3)^{\alpha}{(p)}= 0 ~~~\\
&& {\alpha' {p_\mu} (\tilde \chi_5^3)_{\dot\alpha}{(p)} \bar\sigma_\mu^{\dot\alpha\alpha}} - a_2 (\chi_3^3)^{\alpha}{(p)} - (\chi_6^3)^{\alpha}{(p)}= 0 ~~~\\
&&  {\alpha' {p_\mu} (\chi_6^3)^{\alpha}{(p)}  \sigma^\mu_{\alpha\dot\alpha}} - a_2 (\tilde \chi_4^3)_{\dot\alpha}{(p)} + a_1 (\tilde \chi_5^3)_{\dot\alpha}{(p)} = 0 ~~~
\eea
from the $simple$ poles and for the $double$ pole the usual $\alpha ' p^2={a}_3 $ condition. 
Diagonalizing the system we find
\bea
&& {\cal P} \chi^3_2 = - \tilde\chi^3_1 \\
&& {\cal P}^\dagger \tilde\chi^3_1 = a_3 \chi^3_2 \\
&& {\cal P} \chi^3_3 = - (\tilde\chi^3_4 + \tilde\chi^3_5) \\
&& {\cal P}^\dagger (\tilde\chi^3_4 + \tilde\chi^3_5) = (a_1 + a_2) \chi^3_3 \\
&& {\cal P} \chi^3_6 = a_2 \tilde\chi^3_4 - a_1\tilde\chi^3_5 \\
&& {\cal P}^\dagger (- a_2 \tilde\chi^3_4 + a_1\tilde\chi^3_5) = (a_1 + a_2) \chi^3_6
\eea
where ${\cal P} = \sqrt{\alpha'} p^\mu \bar\sigma_\mu^{\dot\alpha\alpha}$, ${\cal P}^\dagger = \sqrt{\alpha'} {p_\mu} \sigma^\mu_{\alpha\dot\alpha}$ with $ {\cal P} {\cal P}^\dagger = {\cal P}^\dagger {\cal P} = - \alpha' p^2 = \alpha' {m}^2 = a_3 = a_1 + a_2 $.

Therefore we have three Dirac fermions which are BRST invariant:
\bea
\left( \ba{ll} (\chi_2^3)_{[--+]}^{\alpha} \\ (\tilde \chi_1^3)^{[---]}_{\dot\alpha} \ea\right)~,~~~~  
\left( \ba{ll} (\chi_3^3)_{[+++]}^{\alpha} \\ (\tilde\chi^3_4 + \tilde\chi^3_5)^{[\mp\pm+]}_{\dot\alpha} \ea\right) ~,~~~~
\left( \ba{ll} (\chi_6^3)_{[--+]}^{\alpha} \\ (- a_2\tilde\chi^3_4 + a_1\tilde\chi^3_5)^{[\mp\pm+]}_{\dot\alpha} \ea\right) ~,~~~~
\eea
which is what one was expecting. As in previous formulae we have exposed the three R-charges as $[2{r}_1, 2{r}_2, 2{r}_3]:= [a_1+ {r}_1, a_2+ {{r}_2}, - a_3+ {{r}_3}]$ so that $R= {r}_1+{r}_2+{r}_3$ since $a_1+a_2-a_3=0$.

\item Vector with mass $\alpha' {m}^2 = {a}_3$.

Consider the state:
\bea
&& W^\mu ~:~~~ \psi^\mu_{-1/2} |a_1,a_2, a_3 \rangle_{NS}~,~~~~~~~~~~~~
\eea
which satisfies GSO projection. That is a vector multiplet living at the intersection between the two branes. The VO of such state is
\bea
&& V_{W^\mu}
= W_{\mu}(p) \psi^\mu\sigma_{{a}_1 } \sigma_{{a}_2 } \sigma^\dagger_{{a}_3 } 
e^{i[{a}_1 \varphi_1 + {a}_2 \varphi_2 - {a}_3  \varphi_3]} e^{-\varphi} e^{ipX}~~~~
\eea
with mass $\alpha' {m}^2 = {a}_3 $.

Commuting with the BRST charge we get $p{\cdot} W = 0$ from the simple pole. Actually, as mentioned above, one can relax this condition including a non BRST invariant scalar VO that provides the ``longitudinal" mode of the massive vector. 
\end{itemize}

\section{SUSY transformations}

On-shell SUSY transformations act as follows  
\bea
&& [\epsilon Q^{(-1/2)}, V^{(-1/2)}_F(\psi)] = V^{(-1)}_B(\delta_\epsilon\phi)  ~ , \nn\\
&& [\epsilon Q^{(+1/2)}, V^{(-1)}_B(\phi)] = V^{(-1/2)}_F(\delta_\epsilon\psi) = [\epsilon Q^{(-1/2)}, V^{(0)}_B(\phi)] 
\eea
For instance in $D=10$ one has 
\bea
&& [\epsilon Q^{(-1/2)}, V^{(-1/2)}_F(\Lambda) ] = V^{(-1)}_B (A_m = \epsilon \Gamma_m \Lambda) ~ , \nn\\
&& [\epsilon Q^{(+1/2)}, V^{(-1)}_B(A_m)] = V^{(-1/2)}_F(\Lambda = \epsilon \Gamma_{mn} F^{mn}) 
\eea

\subsection{Massless multiplets}

Acting on massless chiral multiplets in $D=4$ one has generically 
\bea
&& [\epsilon Q^{(+3/2,-1/2)}, V^{(-1/2, -1/2)}_{\chi{(k)}}] = V^{(+1, -1)}_{\delta\phi{(p)} = \chi(k)\epsilon}  ~, \quad [\bar\epsilon\bar{Q}^{(-3/2,-1/2)}, V^{(-1/2, -1/2)}_{\chi(k)}] =  0 ~,\nn\\
&& [\epsilon Q^{(+3/2,+1/2)}, V^{(+1, -1)}_{\phi(k)}] = 0 ~ , \quad [\bar\epsilon\bar{Q}^{(-3/2,+1/2)}, V^{(+1, -1)}_{\phi(k)}] =  V^{(-1/2, -1/2)}_{\delta\chi(k) = \sigma^\mu\bar\epsilon k_\mu\phi(k)} ~.
\eea
where the superscripts denote R-charge and super-ghost charge (``picture").

This is what happens acting on the vertex operators 
\bea &&
V^{[001]}_B = \phi(k) \sigma_1 \sigma_2 \sigma_3^\dagger e^{i [a_1\varphi_1 + a_2\varphi_2 + (1- a_3) \varphi_3]} e^{-\varphi} e^{ikX}\nn\\
&& V^{[--+]}_F = \chi^a(k) S_a \sigma_1 \sigma_2 \sigma_3^\dagger e^{ i [(a_1-{1 \over 2})\varphi_1 + (a_2-{1 \over 2})\varphi_2 - (a_3-{1 \over 2})\varphi_3]} e^{-{\varphi \over 2}} e^{ikX}
\eea
forming a massless chiral multiplet with $k^2 = 0$.

\subsection{Massive multiplets}

For massive multiplets the situation is more involved.

\begin{itemize}

\item Massive multiplet with $\alpha' {m}^2 = a_1 = {\theta}_1/\pi$ (the same applies to the massive multiplet with $\alpha' {m}^2 = a_2$).

It consists in a chiral 
\bea
&& V^{[001]}_B = \phi_1(p) \tau_1 \sigma_2 \sigma_3^\dagger e^{ i [a_1\varphi_1 + a_2\varphi_2 + (1- a_3) \varphi_3]}
 e^{-{\varphi}} e^{ipX} , \nn\\
&& V^{[--+]}_F = \chi_1^\alpha(p) S_\alpha \tau_1 \sigma_2 \sigma_3^\dagger e^{ i [(a_1-{1 \over 2})\varphi_1 + (a_2-{1 \over 2})\varphi_2 - (a_3-{1 \over 2})\varphi_3]} e^{-{\varphi \over 2}} e^{ipX}
\eea
and an anti-chiral multiplet 
\bea
&& V^{[0-10]}_B = \tilde\phi^\dagger_1(p) \sigma_1 \sigma_2 \sigma_3^\dagger e^{i [a_1\varphi_1 + (a_2-1)\varphi_2 - a_3 \varphi_3]}  e^{-{\varphi}} e^{ipX}~, \nn\\ 
&& V^{[+-+]}_F = \tilde\chi_1^{\dot\alpha}(p) C_{\dot\alpha}(p) \sigma_1 \sigma_2 \sigma_3^\dagger e^{ i [(a_1+{1 \over 2})\varphi_1 + (a_2-{1 \over 2})\varphi_2 - (a_3-{1 \over 2})\varphi_3]} e^{-{\varphi \over 2}} e^{ipX}
\eea
The SUSY transformations read
\bea
&& [\epsilon{Q}, V^{[--+]}_F(\chi_1(p))] = V_B (\phi^{[001]}(p) = \epsilon\chi_1(p)) \quad , \quad [\bar\epsilon\bar{Q}, V^{[--+]}_F(\chi_1(p))] = 0  ~, \nn\\
&& [\epsilon{Q}, V^{[001]}_B (\phi_1(p))] = 0
~, \\
&& ~~~~~~~~~~ [\bar\epsilon\bar{Q}, V^{[001]}_B (\phi_1(p))] = V^{[--+]}_F (\chi_1 = \sigma^\mu\bar\epsilon p_\mu\phi_1(p)) + V^{[+-+]}_F (\tilde\chi^\dagger_1 = a_1 \phi_1(p) \bar\epsilon) \nn
\eea
and
\bea
&& [\epsilon{Q}, V^{[+-+]}_F(\tilde\chi^\dagger_1(p))]= 0 \quad , \quad [\bar\epsilon\bar{Q}, V^{[+-+]}_F(\tilde\chi^\dagger_1(p))]  =  V^{[0-10]}_B (\tilde\phi_1^\dagger(p) = \bar\epsilon\tilde\chi^\dagger_1(p)) \nn\\
&& [\epsilon{Q}, V^{[0-10]}_B (\tilde\phi^\dagger_1(p))] = V^{[+-+]}_F (\tilde\chi^\dagger_1 = \epsilon\sigma^\mu p_\mu\tilde\phi^\dagger_1(p)) + V^{[--+]}_F (\chi_1 = \tilde\phi^\dagger_1(p)\epsilon) ~, \nn\\  
&& ~~~~~~~~~~ [\bar\epsilon\bar{Q}, V^{[001]}_B (\tilde\phi^\dagger_1(p))] = 0 
\eea
For the (on-shell) fields one has 
\bea
\delta\phi_1(p) = \epsilon \chi_1(p)  \quad , \quad \delta\chi_1(p)  = \sigma^\mu\bar\epsilon p_\mu\phi_1(p) +
\tilde\phi^\dagger_1(p)\epsilon
\eea
 and
\bea
\delta\tilde\phi^\dagger_1(p) = \bar\epsilon \tilde\chi^\dagger_1(p)  \quad , \quad \delta\tilde\chi^\dagger_1(p) = \epsilon \sigma^\mu p_\mu\tilde\phi^\dagger_1(p) +
a_1 \phi_1(p) \bar\epsilon
\eea
Rescaling $\phi_1$ and $\chi_1$ by $1/\sqrt{a_1}$ one gets a more symmetric form where the on-shell F-terms are $F_1(p) = m_1 \tilde\phi^\dagger_1(p)$ and $\tilde{F}^\dagger_1(p) = m_1\phi_1(p)$ with $m_1 = \sqrt{a_1/\alpha'}$, as expected.

\item Massive multiplets with $\alpha' {m}^2 = a_3 = {\theta}_3/\pi$. 

For the fermion vertex operators we find
\bea
&& [\epsilon{Q}, V^{[---]}_F(\bar{u}_1(p))] = V^{[000]}_B (W_\mu(p) = \epsilon\sigma_\mu \bar{u}_1(p)) \quad , \quad \quad [\bar\epsilon\bar{Q}, V^{[---]}_F(\bar{u}_1(p))] = 0  \nn\\
&& [\epsilon{Q}, V^{[--+]}_F ({u}_2(p))] = V^{[001]}_B (\phi_1 = \epsilon{u}_2(p)) \quad , \quad \quad [\bar\epsilon\bar{Q}, V^{[--+]}_F ({u}_2(p))] = 0  \nn\\
&& [\epsilon{Q}, V^{[+++]}_F(u_3(p))] = 0 \quad , \quad \quad [\bar\epsilon\bar{Q}, V^{[+++]}_F(u_3(p))] = V^{[000]}_B (W_\mu(p) = u_3(p)\sigma_\mu \bar\epsilon) \nn\\
&& [\epsilon{Q}, V^{[-++]}_F(\bar{u}_4(p))] = 0 \quad , \quad \quad [\bar{Q}(\bar\epsilon), V^{[-++]}_F (\bar{u}_4(p))] = V^{[-100]}_B (\tilde\phi_3 = \bar\epsilon \bar{u}_4(p)) \nn\\
&& [\epsilon{Q}, V^{[+-+]}_F(\bar{u}_5(p))] = 0 \quad , \quad \quad [\bar{Q}(\bar\epsilon), V^{[+-+]}_F(\bar{u}_5(p))] = V^{[0-10]}_B (\tilde\phi_4 = \bar\epsilon \bar{u}_4(p)) \nn\\
&& [\epsilon{Q}, V^{[--+]}_F (u_6(p))] = V^{[001]}_B (\phi_2 = \epsilon u_6(p)) \quad, \quad \quad  [\bar\epsilon\bar{Q}, V^{[--+]}_F (\chi_6)] = 0
\eea
For the boson vertex operators we find instead
\bea
&& [\epsilon{Q}, V^{[000]}_B (W(p))] = V^{[+++]}_F (\chi_3 = \epsilon\sigma^{\mu}\bar\sigma^{\nu}p_\mu W_\nu(p)) + V^{[+-+]}_F (\tilde\chi_4 = \epsilon\sigma^{\mu}W_\mu(p)) \nn\\
&& ~~~~~~~~~~~~~~~~~~~~~~~~~~~~~~~~~~~~~~~~~~~~~~~~~~~~~~~~~~~~~~~~+ V^{[-++]}_F (\tilde\chi_5 = - \epsilon\sigma^{\mu}W_\mu(p)) ~, \nn\\
&&[\bar\epsilon\bar{Q}, V^{[000]}_B (W(p))] =  V^{[---]}_F (\tilde\chi_1 = \bar\epsilon\bar\sigma^{\mu}\sigma^{\nu} p_\mu W_\nu(p)) + V^{[--+]}_F (\chi_2 = \sigma^{\mu}\bar\epsilon W_\mu(p)) ~, \nn\\
&& [\epsilon{Q}, V^{[001]}_B (\phi_1(p))] = V^{[+++]}_F (\chi_3 = a_3 \epsilon\phi_1(p))~, 
\nn\\
&& [\bar\epsilon\bar{Q}, V^{[001]}_B (\phi_1(p))] = V^{[--+]}_F (\chi_2 = \sigma^\mu\bar\epsilon p_\mu\phi_1(p)) ~, \nn\\
&& [\epsilon{Q}, V^{[001]}_B (\phi_2(p))] = 0~, \nn\\  
&& [\bar\epsilon\bar{Q}, V^{[001]}_B (\phi_2(p))] = V^{[--+]}_F (\chi_6 = \sigma^\mu\bar\epsilon p_\mu\phi_2(p)) + V^{[+-+]}_F (\tilde\chi_4 = a_2 \bar\epsilon p_\mu\phi_2(p)) \nn\\
&&  ~~~~~~~~~~~~~~~~~~~~~~~~~~~~~~~~~~~~~~~~~~~~~~~~~~~~~~~~~~~~ + V^{[-++]}_F (\tilde\chi_5 = a_1 \bar\epsilon p_\mu\phi_2(p)) ~, \nn\\
&&[\epsilon{Q}, V^{[-100]}_B (\tilde\phi_3(p))] = V^{[+-+]}_F (\tilde\chi_4 = \epsilon \sigma^\mu p_\mu\tilde\phi_3(p)) + V^{[--+]}_F (\chi_6 = \epsilon\tilde\phi_3(p))
\quad, \nn\\
&&  [\bar\epsilon\bar{Q}~ , 
V^{[-100]}_B (\tilde\phi_3(p))] = V^{[---]}_F (\tilde\chi_1 = a_1\bar\epsilon \tilde\phi_3(p)) ~, \nn\\
&&[\epsilon{Q}, V^{[0-10]}_B (\tilde\phi_4(p))] = V^{[-++]}_F (\tilde\chi_5 = \epsilon \sigma^\mu p_\mu\tilde\phi_3(p)) + V^{[--+]}_F (\chi_6 = \epsilon\tilde\phi_4(p))
\quad, \nn\\
&&  [\bar\epsilon\bar{Q}~, V^{[-100]}_B (\tilde\phi_3(p))] = V^{[---]}_F (\tilde\chi_1 = a_2\bar\epsilon \tilde\phi_4(p)) ~.
\eea
From these we read the following SUSY transformations for the on-shell fields
\bea
&& \delta W_\mu(p) = \epsilon\sigma_\mu\tilde\chi_1(p) + \bar\epsilon\bar\sigma_\mu\chi_3(p) \\
&& \delta\phi_1(p) = \epsilon\chi_2(p) \\
&& \delta\phi_2(p) = \epsilon\chi_6(p) \\
&& \delta\tilde\phi_3(p) = \bar\epsilon\tilde\chi_4(p) \\
&& \delta\tilde\phi_4(p) = \bar\epsilon\tilde\chi_5(p) \\
&& \delta\tilde\chi_1(p) = W^{\mu\nu}(p) \bar\sigma_{\mu\nu} \bar\epsilon + (a_1 \tilde\phi_3(p) - a_2 \tilde\phi_4(p) + p{\cdot}W(p)) \bar\epsilon  \\
&& \delta\chi_2(p) = \sigma_{\mu}\bar\epsilon (W^{\mu}(p) + p^\mu \phi_1(p))  \\
&& \delta\chi_3(p) = W^{\mu\nu}(p) \sigma_{\mu\nu} \epsilon + (a_3 \phi_1(p) + p{\cdot}W(p)) \epsilon  \\
&& \delta\tilde\chi_4(p) = \epsilon\sigma_{\mu} (W^{\mu}(p) + p^\mu \tilde\phi_3(p)) + a_2 \phi_2(p) \bar\epsilon \\
&& \delta\tilde\chi_5(p)= \epsilon\sigma_{\mu} (-W^{\mu}(p) + p^\mu \tilde\phi_4(p)) + a_1 \phi_2(p) \bar\epsilon \\
&& \delta\chi_6(p) = \sigma_{\mu}\bar\epsilon p^\mu \phi_2(p) + (\tilde\phi_3(p) + \tilde\phi_4(p)) \epsilon 
\eea
Combining $\delta\tilde\chi_4(p)$ and $\delta\tilde\chi_5(p)$ one finds
\bea
\delta(\tilde\chi_4(p) + \tilde\chi_5(p))  = \epsilon\sigma_{\mu}  p^\mu (\tilde\phi_3(p) + \tilde\phi_4(p)) + (a_2 +a_1)\phi_2(p) \bar\epsilon
\eea
and 
\bea
\delta(a_1\tilde\chi_4(p) -a_2 \tilde\chi_5(p))  = \epsilon\sigma_{\mu}  p^\mu (a_1\tilde\phi_3(p) - a_2 \tilde\phi_4(p)) + (a_2 +a_1)\epsilon\sigma_{\mu}  W^\mu(p)~~~~~~
\eea
that allow to disentangle a ``half hyper-multiplet'' (one chiral + one anti-chiral) consisting of $\{\chi_6, \phi_2\}$ and $\{\tilde\chi_4 + \tilde\chi_5, \phi_3 + \phi_4\}$  from a vector multiplet consisting of a vector boson $W_\mu$, two Left-Handed fermions ($\chi_2$ and $\chi_3$), two scalars ($\phi_1$ and $a_1 \phi_3 - a_2 \phi_4$, a combination thereof is eaten by the vector) and two Right-Handed fermions ($\tilde\chi_1$ and $a_1 \tilde\chi_4 - a_2 \tilde\chi_5$).
 
\end{itemize}

\section{Interactions}

In addition to the ``standard" interactions $\ell\ell\ell$ (where $\ell$ stands for light or actually massless) one can consider $H\ell\ell$ ($H$ stands for ``heavy" or massive), $HH\ell$ or $HHH$. 

For simplicity let us focus on the decay process $H\rightarrow \ell_1\ell_2$ with $H$ a boson and $\ell_{1,2}$ two massless fermions. If the fermions have the same chirality then $H$ must be a longitudinal vector ($h=0$ in the CM = rest frame) or a scalar. If the fermions have opposite chirality then $H$ must be a transverse vector boson ($h=\pm 1$ in the CM = rest frame).

\subsection*{$W^\mu\to f_L f_R$}

\begin{figure}[h]
\begin{center}
\epsfig{file=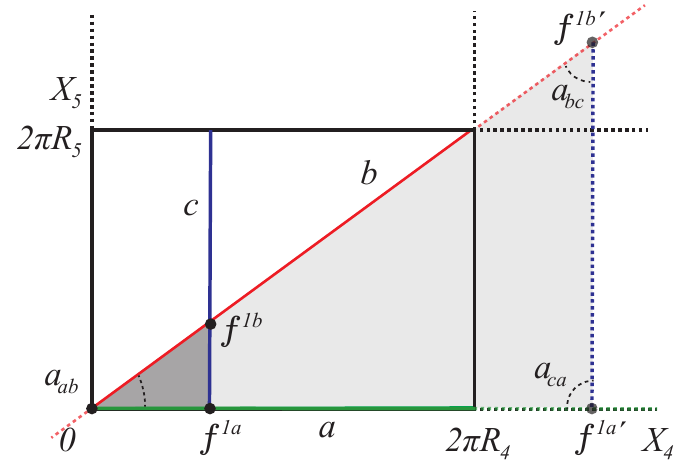,width=80mm}
\caption{We present the first torus with the three intersecting branes $a$, $b$ and $c$. The $f^{1a}$ ($f^{1b}$) denote the intersection points of the $a$ ($b$) brane with the $c$ brane. The $f^{1a'}$ ($f^{1b'}$)... denote the intersections after one wrap of the $a$ ($b$) brane around the $X_4$ ($X_5$) direction. For simplicity we assume that D-branes intersect only once in the fundamental torus.}\label{fig. triangles}
\end{center}
\end{figure}

Let us first consider the case of a massive vector boson. The relevant decay amplitude is
\bea
{\cal A}(W\rightarrow f^L_1f^R_2) &=& \langle cV^{-1/2}_{f_L}(k_1)  cV^{-1/2}_{f_R}(k_2) cV^{-1}_W(p)  \rangle T_{CP}\\
&=&  W^\mu(p)  \bar{v}_{\dot{\alpha}}(k_2) \bar\sigma_\mu^{\dot\alpha\alpha} u_\alpha(k_1) z^{1- {1\over 4} + 2\alpha' k_1k_2}_{12}z^{1-1+ 2\alpha' k_1p}_{13}z^{1-1+ 2\alpha' k_2p}_{23}  \nn\\
&& ~~~ \times \Big\langle \sigma_{a^{ab}_1} \sigma_{a^{ab}_2} \sigma^\dagger_{a^{ab}_{3}}e^{i [(a^{ab}_1-{1\over 2}) \varphi_1 +(a^{ab}_2-{1\over 2}) \varphi_2 - (a^{ab}_3 - {1\over 2}) \varphi_3]} \nn\\
&& ~~~~~ \times \sigma^\dagger_{a^{bc}_{1}} \sigma^\dagger_{a^{bc}_{2}} \sigma_{a^{bc}_3}e^{i [-(a^{bc}_1-{1\over 2}) \varphi_1-(a^{bc}_2-{1\over 2}) \varphi_2 + (a^{bc}_3- {1\over 2}) \varphi_3]}  \nn\\
&& ~~~~~ \times  \sigma_{a^{ca}_1} \sigma_{a^{ca}_2}\sigma^\dagger_{a^{ca}_{3}}e^{i [a^{ca}_1\varphi_1+ a^{ca}_2 \varphi_2 - a^{ca}_3 \varphi_3]} \Big\rangle T_{CP}\nn
\eea
where $2\alpha' k_1k_2 = \alpha'(k_1+k_2)^2 = \alpha' p^2 = - \alpha' {m}^2 = - a^{ca}_3$ while $2\alpha' k_1p = 2\alpha' k_2p = - \alpha' p^2 = \alpha' {m}^2 = a^{ca}_3 $ and the Chan-Paton factor $T_{CP}$ is given by
$$
T_{CP} = \sum_{i_a=1}^{N_a} \sum_{j_b=1}^{N_b}\sum_{k_c=1}^{N_c}T_{ab}^{i_a}{}_{j_b} T_{bc}^{j_b}{}_{k_c} T_{ca}^{k_c}{}_{i_a}
$$ 
for strings in the bi-fundamentals $({\bf N_a}, {\bf \bar{N}_b})$ ($f_L$), $({\bf N_b}, {\bf \bar{N}_c})$ ($f_R$) and $({\bf N_c}, {\bf \bar{N}_a})$ ($W$). 
The (un-excited) twist-field correlator yields
\bea
{\cal C}_{W00}(1,2,3) z^{- {3\over 4} + a^{ca}_3 }_{12}z^{- a^{ca}_3}_{13}z^{- a^{ca}_3}_{23}. 
\eea
where ${\cal C}_{W00}(1,2,3)$ depends on the intersection angles and is given by \cite{Cvetic:2003ch, Lust:2004cx,Cvetic:2007ku}
\bea
{\cal C}_{W00}(1,2,3)  &=& (2\pi)^{3/4}  \Big( \prod_{I=1,2}\Gamma \big[1-a_I^{ab}, a_I^{bc}, 1-a_I^{ca} \big]~
			\Gamma \big[a_3^{ab}, 1 - a_3^{bc}, a_3^{ca} \big]\Big)^{1/4} \sum_{n_I} {e^{-S_{cl}[n_I]}}\nn\\
\eea
with
\bea
\Gamma[a,b,c] = {\Gamma(a) \Gamma(b)\Gamma(c) \over \Gamma(1-a) \Gamma(1-b)\Gamma(1-c)} ~~ .
\eea 
The generic expression for the world-sheet instanton contribution for a single two-torus has been computed in \cite{Cremades:2003qj, Cvetic:2003ch, Abel:2003vv} and takes the form
\begin{align}
\label{eq WS 1}
S_{(I)}[n_I] &=  \frac{1}{ 2 \pi \alpha'}  
\frac{\sin\pi a_I^{bc} \sin\pi  a_I^{ac}}{\sin\pi  a_I^{ab}}  |f^{(1_a)}_I - f^{(1_b)}_I+ n_I {L}^I_c|^2
\end{align}
for fixed $I=1,2,3$. For our specific setup one has to combine the contributions for three tori $T_1^2\times T_2^2\times T_3^2$.
The $f^{(1_a)}_I - f^{(1_b)}_I$ denotes the distance along brane $c$ on torus $T^2_{(I)}$ between the two nearest intersections $f^{(1_a)}$ and $f^{(1_b)}$ of brane $c$ of total length $L_c$ with the two branes $a$ and $b$ (see fig \ref{fig. triangles}). 


In particular one needs
\bea
\label{condWff}
a^{ab}_I - a^{bc}_I + a^{ca}_I = 0 \quad {\rm for~all} \quad I=1,2,3
\eea
The $z$-dependent factors precisely cancel and the rest produces  
\bea
{\cal A}(W\rightarrow f^L_1f^R_2) = {\cal C}_{W00}(1,2,3) W^\mu(p) \bar{v}_{\dot{\alpha}}(k_2) \bar\sigma_\mu^{\dot\alpha\alpha} u_\alpha(k_1)  ~~ .
\eea

In the spinor helicity basis \cite{Bianchi:2016tju,Bianchi:2015lnw,Bianchi:2015yta} (see also  \cite{Lust:2008qc,Anchordoqui:2009mm,Lust:2009pz,Anchordoqui:2009ja}), setting $p=q_1 + q_2 = w_1\bar{w}_1 + w_2\bar{w}_2$, the three independent massive transverse polarisations are $W_+ = w_1 \bar{w}_2$, $W_0 = w_1\bar{w}_1 - w_2\bar{w}_2 = q_1 - q_2$ and  $W_- = w_2 \bar{w}_1$. Up to ${\cal C}_{W00}(1,2,3)$ the corresponding amplitudes are 
\bea
&&{\cal A}(W_0 f^L_1f^R_2) = w_1 u_1 \bar{w}_1 \bar{v}_2 -  w_2 u_1 \bar{w}_2 \bar{v}_2  ~, \nn\\
&&{\cal A}(W_+ f^L_1f^R_2) = w_1 u_1 \bar{w}_2 \bar{v}_2  ~,\quad
{\cal A}(W_- f^L_1f^R_2) = w_2 u_1 \bar{w}_1 \bar{v}_2  ~.
\label{dangerous}
\eea

 
 The presence of a light ``charged" vector boson may be problematic for proton decay or even worse since one expects the same representations of the (fermions of the) Standard Model to be replicated at $\alpha'{m}^2 = a_1, a_2, a_3=a_1 +a_2, 2a_1 + a_2 = a_1 + a_3, a_1 + 2a_2 = a_2 + a_3, \ldots$. One way out -- or at least a way to postpone the problem\footnote{A careful analysis of proton decay or other processes with Baryon or Lepton number violation should be performed which is beyond the scope of the present investigation. For recent work on B and/or L violation in un-oriented D-brane settings, see for instance
\cite{Addazi:2014ila, Addazi:2015rwa, Addazi:2015hka, Addazi:2015yna} and references therein.} -- would be to have $a_1\ll a_2 \approx a_3 = a_2 +a_1$. In this case the only ``light" excitations would be those with mass $a_1, 2a_1, 3a_1, \ldots$ and these do not produce vector bosons but only scalars and fermions, as coded in the helicity super-trace computed earlier. 
Another possibility -- as we will see later -- is to choose specific values for the intersecting angles that eliminate \eqref{dangerous}.


\subsection*{$\phi_0\to f_1^L f_2^R$ and $\phi_1\to f_1^L f_2^R$}

In this section we will evaluate the Yukawa coupling between a massless and a massive scalar and two fermions.
%

The relevant disk amplitude for the massless scalar is 
\bea
{\cal A}(\phi_0 \rightarrow f^L_1f^L_2) &=& \langle cV^{-1/2}_{f_L}(k_1)  cV^{-1/2}_{f_L}(k_2) cV^{-1}_{\phi_0}(k)  \rangle T_{CP} \nonumber\\
&=& \phi_0(p)  u_2^{\alpha}(k_2) u_{1\alpha}(k_1) z_{12}^{1- {1\over 2} + 2\alpha' k_1k_2}z_{13}^{1-{1\over 2}+ 2\alpha' k_1p}z_{23}^{1-{1\over 2}+ 2\alpha' k_2p}  \nn\\
&& ~~~  \times \Big\langle \sigma_{a^{ab}_1} \sigma_{a^{ab}_2} \sigma^\dagger_{a^{ab}_{3}}e^{i [(a^{ab}_1-{1\over 2}) \varphi_1 +(a^{ab}_2-{1\over 2}) \varphi_2 - (a^{ab}_3 - {1\over 2}) \varphi_3]} \nn\\
&&  ~~~~~ \times \sigma_{a^{bc}_{1}} \sigma_{a^{bc}_{2}} \sigma^\dagger_{a^{bc}_{3}}e^{i [(a^{bc}_1-{1\over 2}) \varphi_1+(a^{bc}_2-{1\over 2}) \varphi_2 - (a^{bc}_3- {1\over 2}) \varphi_3]}  \nn\\
&&  ~~~~~ \times  \sigma_{a^{ca}_1} \sigma_{a^{ca}_2}\sigma^\dagger_{a^{ca}_{3}}e^{i [a^{ca}_1\varphi_1+ a^{ca}_2 \varphi_2 + (1- a^{ca}_3) \varphi_3]} \Big\rangle T_{CP}\nn
\eea
where $2\alpha' k_1k_2 = \alpha'(k_1+k_2)^2 = \alpha' k^2 = 0$ while $2\alpha' k_1k = 2\alpha' k_2k =0$ and $T_{CP}$ is the Chan-Paton factor defined as above. For non-vanishing amplitude one needs
\bea
\label{condyuk}
a^{ab}_I + a^{bc}_I + a^{ca}_I = 1 \quad {\rm for} \quad I=1,2 \quad\quad {\rm and} \quad\quad
a^{ab}_3 + a^{bc}_3 + a^{ca}_3 = 2
\eea
giving
\bea
{\cal A}(\phi_0 \rightarrow f^L_1f^L_2) &=&  {\cal Y}_{000}(1,2,3) \phi_0(p) u_2^\alpha(k_2) u_\alpha(k_1)  ~~ ,
\label{Yukawa000}
\eea
and for ``large'' radii, {\it i.e.} suppressing further world-sheet instantons, ${\cal Y}_{000}(1,2,3)$  is given by \cite{Cvetic:2003ch, Lust:2004cx,Cvetic:2007ku}
\bea
{\cal Y}_{000}(1,2,3)  &=& (2\pi)^{3/4} \Big( \prod_{I=1,2} \Gamma[1-a_I^{ab}, 1- a_I^{bc}, 1- a_I^{ca}]~
\Gamma[a_3^{ab}, a_3^{bc}, a_3^{ca}]\Big)^{1/4} 
\eea
which is identical to ${\cal C}_{W00}(1,2,3) $ with the substitution $a^{bc}_i\to 1-a^{bc}_i$ since $f_R\to f_L$.
Notice also that the conditions \eqref{condWff} and \eqref{condyuk} are different therefore we cannot have simultaneously present the Yukawas \eqref{Yukawa000} and the phenomenologically dangerous couplings \eqref{dangerous}\footnote{The condition to satisfy both \eqref{condWff} and \eqref{condyuk} requires $a_3^{bc}=1$ therefore branes $c$ and $b$ should be parallel in the 3rd torus, violating one of our initial assumptions \eqref{angleassumtion}.}.

Next, one can consider the phenomenologically viable case of the decay of a not-so-heavy Higgs, that is a massive copy with $\alpha' m_{\tilde{H}}^2 = a_1\ll 1$ of the Standard Model Higgs with $\alpha' m_H^2 \approx 0$, into two ``massless''  fermions. Later on we will comment on the decay of the replica Higgs into two photons. 
The relevant disk amplitude for the lowest massive scalar of mass $\alpha' {m}^2 = a^{ca}_1$ ($k=1$ case) is 
\bea
{\cal A}(\phi_{1} \rightarrow f^L_1f^L_2) &=& \langle cV^{-1/2}_{f_L}(k_1)  cV^{-1/2}_{f_L}(k_2) cV^{-1}_{\phi_1}(p)  \rangle T_{CP} \nonumber\\
&=& \phi_{1}(p)  u_2^{\alpha}(k_2) u_{1\alpha}(k_1) z_{12}^{1- {3\over 4} + 2\alpha' k_1k_2}z_{13}^{1-{1\over 2}+ 2\alpha' k_1p}z_{23}^{1-{1\over 2}+ 2\alpha' k_2p}  \nn\\
&& ~~~ \times \Big\langle \sigma_{a^{ab}_1} \sigma_{a^{ab}_2} \sigma^\dagger_{a^{ab}_{3}}e^{i [(a^{ab}_1-{1\over 2}) \varphi_1 +(a^{ab}_2-{1\over 2}) \varphi_2 - (a^{ab}_3 - {1\over 2}) \varphi_3]} \nn\\
&& ~~~~~ \times \sigma_{a^{bc}_{1}} \sigma_{a^{bc}_{2}} \sigma^\dagger_{a^{bc}_{3}}e^{i [(a^{bc}_1-{1\over 2}) \varphi_1+(a^{bc}_2-{1\over 2}) \varphi_2 - (a^{bc}_3- {1\over 2}) \varphi_3]}  \nn\\
&& ~~~~~ \times  \tau_{a^{ca}_1} \sigma_{a^{ca}_2}\sigma^\dagger_{a^{ca}_{3}}e^{i [a^{ca}_1\varphi_1+ a^{ca}_2 \varphi_2 + (1- a^{ca}_3) \varphi_3]} \Big\rangle T_{CP}
\eea
where $2\alpha' k_1k_2 = \alpha'(k_1+k_2)^2 = \alpha' p^2 = - \alpha' {m}^2 = - a^{ca}_1$ while $2\alpha' k_1p = 2\alpha' k_2p = - \alpha' p^2 = \alpha' {m}^2 = a^{ca}_1 $. For non-vanishing amplitude one needs again \eqref{condyuk} giving
\bea
{\cal A}(\phi_{1} \rightarrow f^L_1f^L_2) &=& {\cal Y}_{100}(1,2,3) \phi_{1}(p) u_2^\alpha(k_2) u_\alpha(k_1) 
\eea
The ${\cal Y}_{100}(1,2,3)$ is the Yukawa coupling with one excited twist field in the ``first'' direction. For ``large'' radii, {\it i.e.} suppressing further world-sheet instantons, ${\cal Y}_{100}(1,2,3)$  is given by \cite{Cvetic:2003ch, Lust:2004cx,Cvetic:2007ku} \bea
{\cal Y}_{100}(1,2,3)  &=& (2\pi)^{3/4} v^{(1)} e^{-{1\over 2} |v^{(1)}|^2} \Gamma[1-a_1^{ab}, 1- a_1^{bc}, 1-a_1^{ca}] \\
&&\times\Big( \prod_{I=1,2} \Gamma[1-a_I^{ab}, 1- a_I^{bc}, 1- a_I^{ca}]~
\Gamma[a_3^{ab}, a_3^{bc}, a_3^{ca}]\Big)^{1/4}\nn
\eea
with $v^{(1)}_1= f^{(1_a)}_1 - f^{(1_b)}_1/L_c$ denoting the distance along brane $c$ between the two nearest intersections $f^{(1_a)}_1$ and $f^{(1_b)}$ of brane $c$ of total length $L_c$ with the two branes $a$ and $b$, at whose intersection the light massive string state is localised. Notice that ${\cal Y}_{100}(1,2,3)=0$ when the three stacks of branes intersect at the same point. The maximum $1/\sqrt{e}$ of the function $v e^{-{1\over 2} |v|^2}$ is reached at $v = 1$. 


Taking into account the mass $m_f$ the Standard Model fermions acquire after the Brout-Englert-Higgs mechanism, the decay width of the ``first'' replica Higgs $H_1$ into a fermion pair reads
\be
\Gamma(H_1\rightarrow f\bar{f}) = {N_c(f) m_{H_1} |{\cal Y}_{100}|^2 \over 8\pi} 
\left( 1 - {4 m_f^2 \over m_{H_1}^2}\right)^{3/2}  \ee
where $N_c(f) =3 $ for quarks and $N_c(f)=1$ for leptons. For large extra dimensions and small $a_1^{ca}$, one can take into account only the first term in the instanton sums and approximately find
\be
{\cal Y}_{100} \approx  {\cal Y}_{000}   { v^{(1)} \sin\pi a_1^{ca} \over e^{{1\over 2} |v^{(1)}|^2}\pi a_1^{ca} } = {m_f v^{(1)} \sin\pi a_1^{ca} \over v_H e^{{1\over 2} |v^{(1)}|^2}\pi a_1^{ca} } < {m_f \over v_H \sqrt{e}} 
\ee
where $v_H = 246 GeV$. For small angles $\sin\pi a_1^{ca} / \pi a_1^{ca} \approx 1$ and the partial width is roughly $m_{H_1}/ e m_{H_0} = 750 / e 125 \approx 2.207$ larger than the partial width of the SM Higgs. 

The next replica $H_2$ is expected to have a mass 
\be
m_{H_2} = 1053 GeV = \sqrt{ 2\times750^2 - 125^2} = \sqrt{ 125^2 + 2\times(750^2 - 125^2) }
\ee
and can decay into massless, massive and ``mixed'' pairs of fermions or vector bosons at tree level.

While the decay channel of (replica) Higgses into a fermion pair is allowed at tree level, as the above computation on the disk shows, the decay channel into two photons seems forbidden. For instance for the Standard Model Higgs it requires a loop of massive particles (top quark $t$ or $W$ bosons). Yet in string theory the effective operator
\bea
{\cal L}_{eff} = \alpha' H H^\dagger F_{\mu\nu} F^{\mu\nu}
\eea
or better (in supersymmetric cases ) 
\bea
{\cal L}_{eff} = \alpha' [ {\rm Re}(H_u H_d)  F_{\mu\nu} F^{\mu\nu} + {\rm Im}(H_u H_d)  F_{\mu\nu} \tilde{F}^{\mu\nu}] ~~,
\eea
which are gauge invariant and vanish in the naive field theory limit $\alpha' \rightarrow 0$, may be generated by tree-level exchange of massive open string states (Regge recurrences). Indeed, although the leading field-theory term cancels for any consistent choice of the hyper-charge embedding, sub-leading terms suppressed by $\alpha'$ survive. Once the Higgs field gets a VEV, ${\cal L}_{eff}$ can trigger Higgs decays into two photons. This applies also to a light massive Higgs with $m_{\tilde{H}}\approx 750~GeV$ and the same Chan-Paton charges as the ``massless'' one. 

\begin{figure}[h]
\begin{center}
\epsfig{file=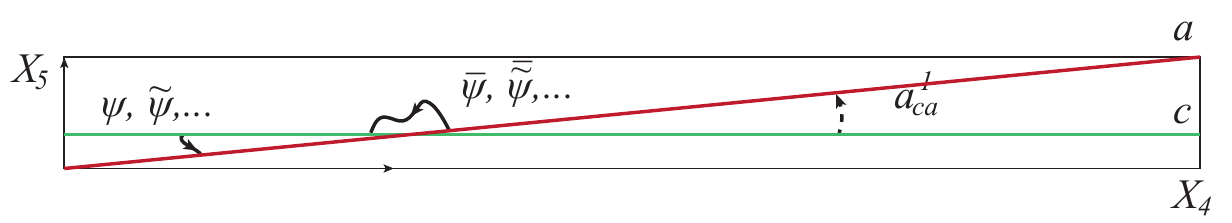,width=120mm}
\caption{The first torus with large $Z_4$ direction. The angle $a_{ca}^1$ becomes very small.  
}\label{fig. TorusWithZ4large}
\end{center}
\end{figure}

The detailed computation of the decay rate into two photons is beyond the scope of the present investigation very much as the embedding of the Standard Model into a ``locally'' or even ``globally'' consistent compactifications with open and unoriented strings. For recent work on how to connect String Theory to Particle Physics and Cosmology see for instance 
\cite{Honecker:2015qba} and references therein.
For our interpretation to work, anistropic compactifications with one direction parametrically smaller with respect to the other ones are preferred, see fig \ref{fig. TorusWithZ4large}.

The string picture advocated here for di-photon excess(es) is different from the ones suggested by \cite{Anchordoqui:2015jxc} or \cite{Cvetic:2015vit} where the low massive string state  is either a neutral closed string ``modulus" (dilaton-like) appearing in the gauge-kinetic function of the photon or an exotic state.


\section{Conclusions and outlook}

We have further investigated light massive string states that appear at D-brane intersections \cite{Anastasopoulos:2011hj, Anastasopoulos:2011gn, Anastasopoulos:2014lpa, Anastasopoulos:2015dqa}. For small angles they can be parametrically lighter than standard Regge excitations. We have carefully computed the character-valued partition function (helicity super-trace) and identified the relevant physical states. We have then specialised to the case of supersymmetric intersections, determined the BRST invariant vertex operators and reconstructed the super-multiplet structure. Finally, we have computed the decay rate of a massive scalar or vector into two massless fermions, and suggested an alternative interpretation of the 750 GeV diphoton excess at LHC in terms of light string states. We have argued that an effective Higgs coupling to two photons could be generated by integrating out massive string states. Replicas of the (supersymmetric) Standard Model Higgses appear at brane intersections that can have similar couplings. 
We plan to further explore this possibility in the near future by computing various decay rates of light massive string states into massless particles both at tree level and one-loop \cite{AnastasopoulosBianchiGoudelisSengupta}.

\section*{Acknowledgements}

The authors would like to thank Andrea Addazi, Dario Consoli, Francesco Fucito, Andrea Leonardo Guerrieri, Elias Kiritsis, Jos\`e Francisco Morales Morera, Tassos Petkou, Lorenzo Pieri, Gianfranco Pradisi, Fabio Riccioni, Yassen Stanev for useful discussions and special thanks to Robert Richter for enjoyable collaboration on these and related topics.
MB would like to thank TUW IfTP for hospitality during completion of this work. MB was partly supported by the INFN network ST$\&$FI ``String Theory and Fundamental Interactions'' and by the Uncovering Excellence Grant STaI ``String Theory and Inflation''.
PA was supported by the FWF project P26731-N27.

\newpage
\appendix

\section{Theta functions and Riemann identity}\label{Theta functions and Riemann identity}

In this section we present some formulas which are useful for our computations. 
Assuming $1\gg {u}_I  \ge 0$ as well as the ``triangular inequality" $\sum_I {u}_I  \ge 2 {u}_K$ we have 
\bea
&& \vartheta_1(v) = (z^{+1/2} - z^{-1/2}) q^{1/8} \prod_{n=1}^\infty (1-q^n) (1-z q^n) (1-z^{-1} q^n)\\
&& \vartheta_1(u_I) = - q^{1/8-{a}_I /2} (1 - q^{{a}_I }) \prod_{n=1}^\infty (1-q^n) (1- q^{n+{a}_I }) (1- q^{n-{a}_I })
\eea
as well as 
\bea
&&\vartheta_1\left({1\over 2} (v+\epsilon\sum_I u_I)\right) = - \epsilon z^{-\epsilon/4} q^{1/8 - \sum_I {a}_I /4} (1 - z^{\epsilon /2}q^{\sum_I {a}_I /2}) \\
&&~~~~~~~~~~~~~~~~~~~~~~~~~~~~~ \times \prod_{n=1}^\infty (1-q^n) (1-z^{\epsilon /2} q^{n+\sum_I {a}_I /2}) (1-z^{-\epsilon /2} q^{n-\sum_I {a}_I /2}) \nn\\
&&\vartheta_1\left({1\over 2} (v+\epsilon\sum_I u_I) - \epsilon u_K\right) = - \epsilon z^{-\epsilon/4} q^{1/8 - \sum_I {a}_I /4 + {a}_K/2} 
(1 - z^{\epsilon/2}q^{\sum_I {a}_I /2- {a}_K}) \\
&&~~~~~~~~~~~~~~~~~~~~~~~~~~~~~  \times \prod_{n=1}^\infty (1-q^n) (1-z^{\epsilon/2} q^{n+\sum_I {a}_I /2 -{a}_K}) 
(1-z^{-\epsilon/2} q^{n-\sum_I {a}_I /2 + {a}_K})\nn
\eea
The Riemann identity for the even spin structures reads 
\bea
&&\sum_{a, {\rm even}} \vartheta_a(v) \vartheta_a(u_1) \vartheta_a(u_2) \vartheta_a(u_3)=  
\sum_{\epsilon=-1,1} \vartheta_1\left({1\over 2} (v+ \epsilon\sum_I u_I)\right) \prod_K \vartheta_1\left({1\over 2} (v-\epsilon\sum_I u_I + \epsilon u_K) \right)\nn\\
\eea

\section{Useful OPE's}

The OPE's of the twisted fields are
\bea
&& Positive~angles\nn\\
&& ~~~ \ba{lllll}
& \partial Z (z) \, \sigma^+_{a} (w) \sim (z-w)^{a-1} \tau^+_{a} (w)       &~~~~~ &  \partial Z^* (z) \, \sigma^+_{a} (w) \sim (z-w)^{-a} \widetilde{\tau}^+_{a} (w) \\ 
& \partial Z (z) \, \tau^+_{a} (w) \sim (z-w)^{a-1} \omega^+_{a} (w)       &&  \partial Z^* (z) \, \tau^+_{a} (w) \sim a(z-w)^{-a-1} \sigma^+_{a} (w)  \\
 & \partial Z (z) \, \omega^+_{a} (w) \sim (z-w)^{a-1} \rho^+_{a} (w)      &&  \partial Z^* (z) \, \omega^+_{a} (w) \sim 2a (z-w)^{-a-1} \tau^+_{a} (w)  \\
& \partial Z (z) \, \widetilde{\tau}^+_{a} (w) \sim (1-a) (z-w)^{-2+a} \sigma^+_{a} (w)  && \partial Z^* (z) \, \widetilde{\tau}^+_{a} (w) \sim (z-w)^{-a} \widetilde{\omega}^+_{a} (w) \\
 & \partial Z (z) \, \widetilde\omega^+_{a} (w) \sim 2(1-a) (z-w)^{-2+a} \widetilde\tau^+_{a} (w)  &&  \partial Z^* (z) \, \widetilde\omega^+_{a} (w) \sim (z-w)^{-a} \widetilde{\rho}^+_{a} (w)\nn 
\ea\nn
\eea
Schematically, we can depict the above OPE's in the following diagram
\bea
\ba{ccccccccccccc} 
&& \tilde \omega^\pm_a && && \sigma^\pm_a && && \omega^\pm_a\\
&& & \nwarrow  && \nearrow  &&  \nwarrow  && \nearrow \\
&& && \tilde \tau^\pm_a  && &&  \tau^\pm_a \\
&& &&  & \nwarrow  && \nearrow \\
&& &&  && \sigma^\pm_a \ea
\eea
where $\nearrow$, $\nwarrow$ denote the action of $\partial Z$, $\partial Z^*$ on various twisted fields and can be easily extended to higher excited bosonic twist fields.

With these OPE's one can determine the conformal dimension of the respective twist fields. We summarize our results in table \ref{table conformal dimensions}.
\begin{table}[h] \centering
\begin{tabular}{| l | l || l | l |}
\hline
Fields & conformal dimensions & Fields & conformal dimensions \\ \hline \hline
$\sigma^+_{a}$    &   $\frac{1}{2}a(1-a)$   &   $\sigma^-_{a}$    &    $\frac{1}{2} a(1-a) $    \\
$\tau^+_{a}$    &   $\frac{1}{2}a(3-a)$   &  $\tau^-_{a}$    &   $\frac{1}{2}( 2+a) (1-a)$  ~~~~~~~~~~~~    \\
$\omega^+_{a}$    &   $\frac{1}{2}a(5-a)$   &      $\omega^-_{a}$   &    $\frac{1}{2}(a+4)(1-a) $  \\
$\widetilde{\tau}^+_{a}$    &   $\frac{1}{2}(a+2)(1-a)~~~$   &   $\widetilde{\tau}^-_{a}$   &    $ \frac{1}{2} a(3-a)$   \\
$\widetilde{\omega}^+_{a}$    &   $\frac{1}{2}(a+4)(1-a)~~~~~~~~~~~~$   &    $\widetilde{\omega}^-_{a}$   &    $\frac{1}{2} a(5-a) $    \\
\hline
\end{tabular}\nn
\caption{\small {The conformal dimensions of bosonic twist fields.}} 
\label{table conformal dimensions}
\end{table}
The above OPE's suggest the following identifications among twist- and anti-twist fields
\begin{align}
\sigma^{-}_{a} (z) = \sigma^+_{1-a} (z) \qquad \qquad \tau^-_{a} (z) = \tau^+_{1-a} (z) \qquad \qquad \widetilde\tau^-_{a} (z) = \widetilde\tau^+_{1-a} (z)
\end{align}  
which can be easily generalized to higher excited twist fields. In addition, conjugate of each field is the tilded one for the oposite angle
\bea
(t^\pm_a)^\dagger = \tilde t^\mp_{a} = \tilde t^\pm_{1-a} 
\eea
for $t=\sigma,~\tau,~\omega$ etc.

We also need the following OPE\footnote{For twist correlators involving higher excited bosonic twist fields, see \cite{David:2000yn,Lust:2004cx, Conlon:2011jq, Pesando:2011ce,Pesando:2012cx, Anastasopoulos:2013sta, Pesando:2014owa}.}:
\bea
&& e^{q_1\varphi(z)} e^{q_2\varphi(w)} \sim (z-w)^{-q_1q_2}  e^{(q_1+ q_2)\varphi(w)}   \\
&& e^{i r_1H(z)} e^{i r_2H(w)} \sim (z-w)^{r_1r_2}  e^{i(r_1+ r_2)H(w)}   \\
&& \psi^\mu(z) \psi^\nu(w) \sim \frac{\eta^{\mu \nu}} {(z-w)}     \\
&& \partial^\mu X(z) e^{ikX(w)} \sim -\frac{2i\alpha' k^\mu}{z-w} e^{ikX(w)}  \\
&& \psi^\mu(z) S_{\a} (w) \sim \frac{1}{\sqrt{2}} \frac{\sigma^\mu_{\alpha\dot\alpha} C^{\dot\alpha}(w)}{(z-w)^{1/2}}  \\
&& \psi_\mu(z) C^{\dot\alpha} (w) \sim \frac{1}{\sqrt{2}} \frac{\bar \sigma_\mu^{\dot\alpha\alpha} S_\alpha(w)}{(z-w)^{1/2}}\\  
&& \psi^\mu(z) S_{\a} (w_1) C_{\dot\alpha}(w_2) \sim \frac{1}{\sqrt{2}} \frac{\sigma^\mu_{\alpha\dot\alpha} }{(z-w_1)^{1/2}(z-w_2)^{1/2}} 
\eea
The three-point bosonic twist field correlator is given by \cite{Cvetic:2003ch, Lust:2004cx,Cvetic:2007ku}\footnote{In case the intersection angles add up to $2$ the correlator takes the form
\bea
\big\langle         \sigma^+_{a_1} (x_1) \sigma^+_{a_2} (x_2) \sigma^+_{a_3} (x_3) \big\rangle =\left( 2 \pi \frac{\Gamma(a_1)\Gamma(a_2)\Gamma(a_3)}{\Gamma(1-a_1)\Gamma(1-a_2)\Gamma(1-a_3)} \right)^{\frac{1}{4}} x^{-(1-a_1) \,(1- a_2)}_{12}\, x^{-(1-a_1) \, (1-a_3)}_{13} \, x^{- (1-a_2) \,(1- a_3)}_{23} \nn\\
\eea
}
\bea
\big\langle         \sigma^+_{a_1} (x_1) \sigma^+_{a_2} (x_2) \sigma^+_{a_3} (x_3) \big\rangle =\left( 2 \pi \frac{\Gamma(1-a_1)\Gamma(1-a_2)\Gamma(1-a_3)}{\Gamma(a_1)\Gamma(a_2)\Gamma(a_3)} \right)^{\frac{1}{4}} x^{-a_1 \, a_2}_{12}\, x^{-a_1 \, a_3}_{13} \, x^{- a_2 \, a_3}_{23} \nn\\ 
\label{eq three point bosonic twist}
\eea

\section{State-Vertex operator dictionary}

In this section we provide a dictionary between states on the worldsheet and their contribution to the VO \cite{Anastasopoulos:2011hj}
\begin{table}[h] \centering
\begin{tabular}{| l | l || l | l |}
\hline
\multicolumn{2}{|c||}{NS sector}&
\multicolumn{2}{|c|}{R sector} 
\\
\hline
\hline
state & vertex operator &state & vertex operator\\ \hline \hline
$| \, a\, \rangle_{NS} $ & 
$  e^{i a H(z)} \sigma^+_{a}(z) $ & $| \, a \, \rangle_{R}  $& 
$  e^{i (\frac{1}{2}-a) H(z)} \sigma^+_{a}(z) $\\ 
\hline
$\alpha_{-a} | \, a\, \rangle_{NS}$ & 
$  e^{i a H_(z) }\tau^+_{a}(z)$ & 
$\alpha_{-a} | \, a\, \rangle_R$ & 
$  e^{i (\frac{1}{2}-a) H(z) }\tau^+_{a}(z) $\\ 
\hline
$\left(\alpha_{-a} \right)^2| \, a\, \rangle_{NS}$ & 
$  e^{i a H(z)} \omega^+_{a}(z) $&
$ \left(\alpha_{-a} \right)^2| \, a\, \rangle_{R}$ & 
$  e^{i (\frac{1}{2}-a) H(z)} \omega^+_{a}(z)$ \\ 
\hline
$ \psi_{-\frac{1}{2} + a}| \, a \, \rangle_{NS} $ & 
$ e^{i \left(a-1\right)H(z)} \sigma^+_{a}(z)$ & 
$\psi_{-a}| \, a \, \rangle_{R} $ & 
$ e^{i (\frac{1}{2}+a) H (z)} \sigma^+_{a}(z) $ \\ 
\hline
$ \alpha_{-a}\, \psi_{-\frac{1}{2} + a} | \, a\, \rangle_{NS}$ & 
$  e^{i \left(a -1\right)H(z) }\tau^+_{a}(z) $ &
$ \alpha_{-a}\, \psi_{-a} | \, a\, \rangle_{R}$ & 
$  e^{i (\frac{1}{2}+a) H(z) }\tau^+_{a}(z) $ \\ 
\hline
$ \left(\alpha_{-a} \right)^2 \psi_{-\frac{1}{2} + a} | \, a\, \rangle_{NS}$ & 
$  e^{i\left(a-1\right) H(z)} \omega^+_{a}(z) $& 
$ \left(\alpha_{-a} \right)^2 \psi_{-a} | \, a\, \rangle_{R}$ & 
$  e^{i(\frac{1}{2}+a) H(z)} \omega^+_{a}(z)  $\\ 
\hline
$ \alpha_{-1+a} | \, a\, \rangle_{NS}$ & 
$  e^{i a H(z) }\widetilde{\tau}^+_{a}(z) $ & 
$\alpha_{-1+a} | \, a\, \rangle_{R}$ & 
$  e^{i (\frac{1}{2}-a) H(z) }{\widetilde\tau}^+_{a}(z) $ \\ 
\hline
$ \alpha_{-1+a} \psi_{-\frac{1}{2} + a} | \, a\, \rangle_{NS}$ &
$  e^{i \left(a -1\right)H(z) } \widetilde{\tau}^+_{a}(z) $ & 
$\alpha_{-1+a} \psi_{-a} | \, a\, \rangle_{R}$ & 
$  e^{i (\frac{1}{2}+a) H(z) } {\widetilde\tau}^+_{a}(z) $\\ 
\hline
\end{tabular}\nn
\caption{\small {Excitations and their corresponding vertex operator part for positive angles.}} 
\label{table Excitations for positive angles}
\end{table}

\clearpage \nocite{*}

\providecommand{\href}[2]{#2}\begingroup\raggedright\endgroup

\end{document}